\documentclass[aps,prx,amsmath,amssymb,reprint,nofootinbib]{revtex4-1}\usepackage{graphicx}
\usepackage{overpic}
\usepackage{epstopdf}
\usepackage{color}
\usepackage{mathtools}
\usepackage[caption=false]{subfig}
\usepackage{empheq}
\definecolor{light-gray}{gray}{0.85}
\newcommand{\shadedbox}[1]{\colorbox{light-gray}{$\displaystyle #1$}}

\newcommand\duetoref[1]{\stackrel{(\ref{#1})}{=}}
\newcommand\duetoreftwo[2]{\stackrel{(\ref{#1})(\ref{#2})}{=}}

\newcommand\evfield{|\Psi\rangle}
\newcommand\helplusmy{|k\ j\ m_y\ +\rangle}
\newcommand\helminusmy{|k\ j\ m_y\ -\rangle}
\newcommand\helmy{|k\ j\ m_y\ \lambda\rangle}
\newcommand{\thi}{\theta_{in}}

\newcommand\zhat{\mathbf{\hat{z}}}
\newcommand\xhat{\mathbf{\hat{x}}}
\newcommand\yhat{\mathbf{\hat{y}}}
\newcommand\shat{\text{TE}}
\newcommand\phat{\text{TM}}
\newcommand\My{M_{\yhat}}
\newcommand\pp{\mathbf{p}}
\newcommand{\Am}{|k\ j\ m_i\ (e)\rangle}
\newcommand{\Ae}{|k\ j\ m_i\ (m)\rangle}
\newcommand{\ajiy}{\beta_{j,m_i}^{(\tau)}}
\newcommand{\ajie}{\beta_{j,m_i}^{(e)}}
\newcommand{\ajim}{\beta_{j,m_i}^{(m)}}
\newcommand{\Ei}{|\Phi_{inc}\rangle}
\newcommand{\Es}{|\Phi_{sca}\rangle}

\newcommand\oj{\overline{j}}
\newcommand\mzbar{\overline{m}_z}
\newcommand\mbar{\overline{m}}
\newcommand\mz{m_z}

\newcommand\lmzbar{\langle \lambda \ \mzbar\ \oj\ k|}
\newcommand\rmz{|k \ j\ \mz\ \lambda\rangle}
\newcommand\rmzo{|k \ j\ \mz\ \lambda\rangle_o}

\begin{document}
\title{Transverse multipolar light-matter couplings in evanescent waves}
\author{Ivan Fernandez-Corbaton}
\email{ivan.fernandez-corbaton@kit.edu}
\affiliation{Institute of Nanotechnology, Karlsruhe Institute of Technology, 76021 Karlsruhe, Germany}

\author{Xavier Zambrana-Puyalto}\email{xavislow@protonmail.com}
\affiliation{Aix Marseille Univ, CNRS, Centrale Marseille, Institut Fresnel, Marseille, France}
\author{Nicolas Bonod}\affiliation{Aix Marseille Univ, CNRS, Centrale Marseille, Institut Fresnel, Marseille, France}
\author{Carsten Rockstuhl}\affiliation{Institut f\"ur Theoretische Festk\"orperphysik, Karlsruhe Institute of Technology, 76131 Karlsruhe, Germany\\
Institute of Nanotechnology, Karlsruhe Institute of Technology, 76021 Karlsruhe, Germany}

\begin{abstract}
	We present an approach to study the interaction between matter and evanescent fields. The approach is based on the decomposition of evanescent plane waves into multipoles of well-defined angular momentum transverse to both decay and propagation directions. We use the approach to identify the origin of the recently observed directional coupling of emitters into guided modes, and of the opposite Zeeman state excitation of atoms near a fiber. We explain how to rigorously quantify both effects, and show that the directionality and the difference in excitation rates grow exponentially with the multipolar order of the light-matter interaction. We also use the approach to study and maximize the transverse torque exerted by an evanescent plane wave onto a given spherical absorbing particle. All the obtained physical insights can be traced back to the two main features of the decomposition of evanescent plane waves into transverse multipolar modes: A polarization independent exponential dominance of modes with large transverse angular momentum, and a polarization controlled parity selection rule.
\end{abstract}
\maketitle

\section{Introduction and summary}
The analysis and engineering of light-matter interactions are central subjects in electrodynamics. The current miniaturization trends in nanophotonics increase the demands on the performance of theoretical models for interactions at close quarters. The interaction between a molecule and the guided mode of a nearby waveguide, or between a quantum dot and a nearby nano-antenna are two examples of current interest. The properties of evanescent fields and the light-matter couplings mediated by them are a key part of any nanophotonics theory. 

The interaction between near fields and matter systems, often atoms, has been the subject of both experimental and theoretical research, e.g: \cite{Mabuchi1994,Van1994,Novotny1996,Tojo2004,Tojo2005,Bharadwaj2007,Lembessis2009,Andrews2010,Le2014,Reitz2014,Bliokh2014b,Mitsch2014,Petersen2014,Rodriguez2014,Sollner2015,Bliokh2015,Van2016}. The research shows that the angular momentum properties of evanescent fields are different from those of propagating fields, and that these differences are important for understanding the interactions. For example, in very recent developments, the concept of a transverse electromagnetic spin has been introduced \cite{Bliokh2012,Bliokh2014b,Bliokh2015}, and used to explain experimental results showing directional light-matter coupling mediated by evanescent fields \cite{Petersen2014,Rodriguez2014,Sollner2015}. In the context of transverse electromagnetic spin, this is known as transverse spin-momentum locking. The concept of transversely spinning fields is also treated in Refs. \onlinecite{Aiello2009,Aiello2015}.

In this article, we combine a complete multipolar decomposition of the evanescent fields with the idea of choosing a transverse axis to quantize the angular momentum. In Sec. \ref{sec:tjy}, we decompose an evanescent plane wave into multipolar modes with well-defined angular momentum around the direction transverse to both decay and propagation directions. The transverse multipolar structure reveals a polarization independent exponential predominance of modes with large multipolar order and large transverse angular momentum. The dominant modes have either large positive or large negative values of the transverse angular momentum. The sign of the dominant transverse angular momentum switches with the propagation direction, i.e. it is locked to it. This exponential trend increases the relevance of higher order multipolar couplings in evanescent light-matter interactions. We also derive a polarization dependent selection rule:  Half of the coefficients in the transverse multipolar decomposition of a TE(TM) plane wave are zero. Given a transverse multipolar order and angular momentum value, the TE plane waves have zero content of one of the two possible multipole parities and the TM plane waves have zero content of the other one. The selection rule applies not only to TE/TM plane waves, but to any pair of modes with opposite eigenvalue under a mirror reflection across the plane perpendicular to the direction of angular momentum quantization. In particular, the selection rule controls the selective interaction between mirror symmetric evanescent fields and the electric or magnetic multipoles of the matter particles immersed in them. The physical consequences of the selection rule and the increased relevance of higher multipolar orders in evanescent fields are clearly identified in all the practical situations that we consider in later sections.

In Sec. \ref{sec:exp}, we use the theory developed in Sec. \ref{sec:tjy} to analyze the recently observed directional coupling of emitters into guided modes \cite{Petersen2014,Rodriguez2014,Sollner2015}, and the simultaneous preparation of different atomic states using the same guided optical mode \cite{Mitsch2014}. We explain how to rigorously quantify both effects, and show that the directionality and the difference in excitation rates grow exponentially with the multipolar order of the light-matter interaction. 

In Sec. \ref{sec:torque} we use the transverse multipolar decomposition developed in Sec. \ref{sec:tjy} to study the transverse torque exerted by an evanescent plane wave onto an absorbing sphere. We first derive the formulas to compute the torque exerted on a sphere by an incident beam expressed as a superposition of transverse multipolar modes. Then, we find the plane wave parameters that maximize the torque exerted on a silicon sphere with a given radius immersed in an evanescent plane wave; that is, the optimal combination of frequency, angle of incidence, and polarization of the corresponding propagating plane wave undergoing total internal reflection. In the example that we consider, the optimal frequency coincides with the magnetic quadrupolar absorption resonance of the sphere. The optimal polarization is found to be TE polarized, which, according to the aforementioned selection rule, and contrary to the TM polarization, can excite the magnetic quadrupolar absorption resonance of the sphere with the largest possible value of transverse angular momentum.

The physical insights obtained in the applications highlight that both the selection rule and the enhanced role of multipolar couplings of higher order are crucial for precisely understanding and engineering the interaction of evanescent light with matter.   

\begin{figure}
	\includegraphics[width=\linewidth]{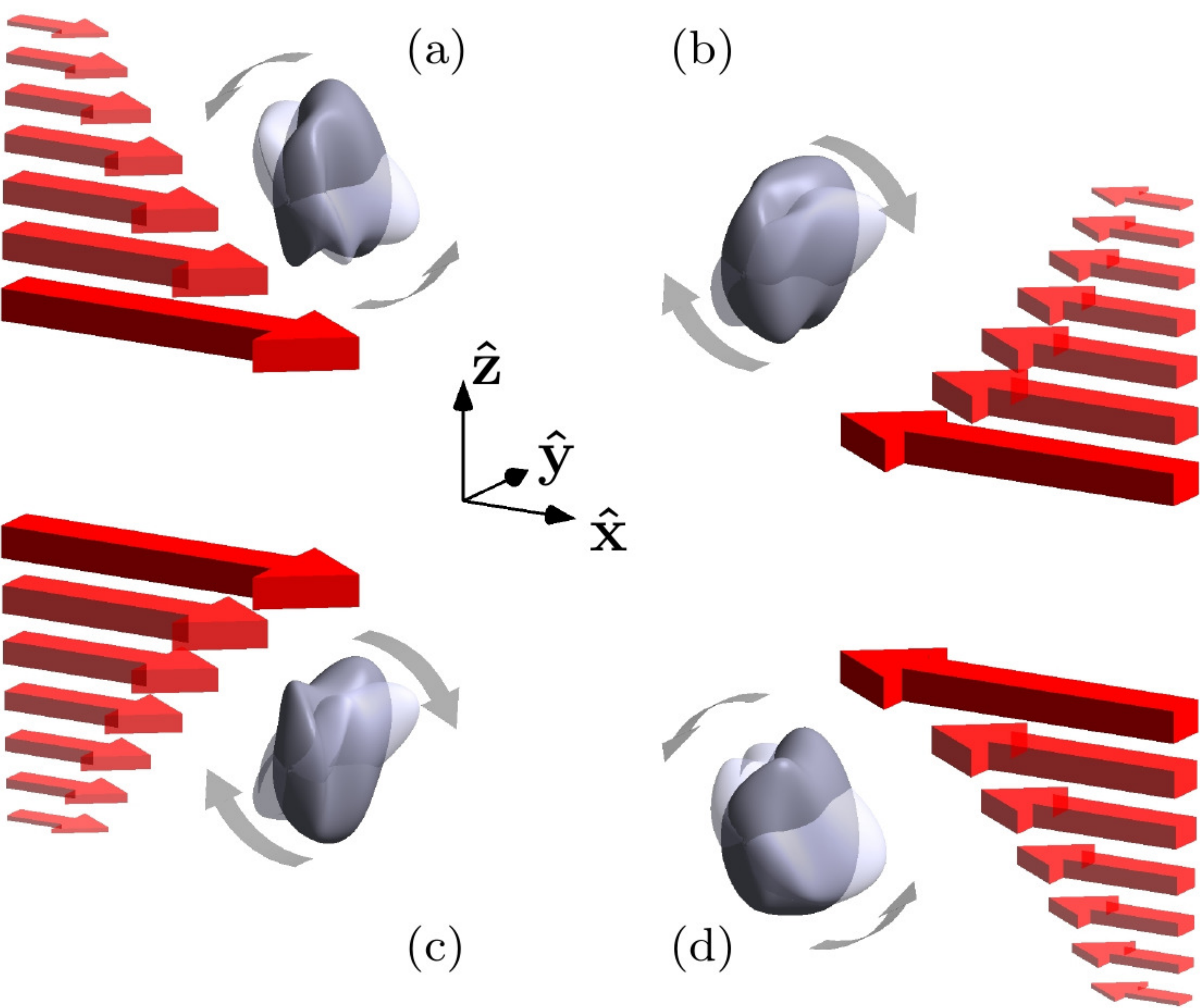}
	\caption{\label{fig:intuition} Objects immersed in electromagnetic fields. The fields decay along a given direction and have a well-defined linear momentum component along a direction orthogonal to the first one. This gives rise to a corresponding Poynting vector component with decaying amplitude, represented by the straight red arrows. The fields induce torques on the objects around a third direction orthogonal to both decay and momentum. The torques are depicted as curved gray arrows. The sense of the torque depends on the decay and momentum directions.}
\end{figure}

\section{Transverse angular momentum analysis of evanescent plane waves\label{sec:tjy}}
\subsection{Transverse angular momentum}
We start by considering the situations shown in Fig. \ref{fig:intuition}(a). An object is illuminated by different electromagnetic fields. The intensity of the fields decreases along a given direction ($+\zhat$ or $-\zhat$ in the figure). Additionally, the field has a well-defined linear momentum component in a direction orthogonal to the first one ($+\xhat$ or $-\xhat$ in the figure). We may imagine that the fields ``push'' the objects as indicated by the straight red arrows. In panel (a) the pushes are strongest near the bottom of the object and decrease in strength as $z$ increases, being weakest near the top of the object. Our intuition tells us that the $z$ dependent pushes will create a torque around the third axis $\yhat$. The torque is represented by the curved gray arrows. We also intuitively understand that the sense of the torque will depend on both the momentum and decay directions.  When the horizontal arrows point along $-\xhat$ instead of $\xhat$, the torque is created in the opposite sense [Fig. \ref{fig:intuition}(b)]. The sense of the torque also reverses when the field decreases along $-\zhat$ [Fig. \ref{fig:intuition}(c)]. The two sign changes compensate each other when flipping both momentum and gradient [Fig. \ref{fig:intuition}(d)]. 

We would now like to turn our intuition about the torque around $\yhat$ into physically relevant quantities. To this end, we propose to expand the electromagnetic field in a basis of modes with well-defined angular momentum around the $\yhat$ direction ($J_y$). This choice is motivated by the conservation of angular momentum. For example, when an object absorbs radiation with a value of $J_y$ equal to $m_y$, it experiences a torque proportional to $m_y$ \cite{Friese1996,Simpson1997}. 

\begin{figure}[h!]
	\includegraphics[width=\linewidth]{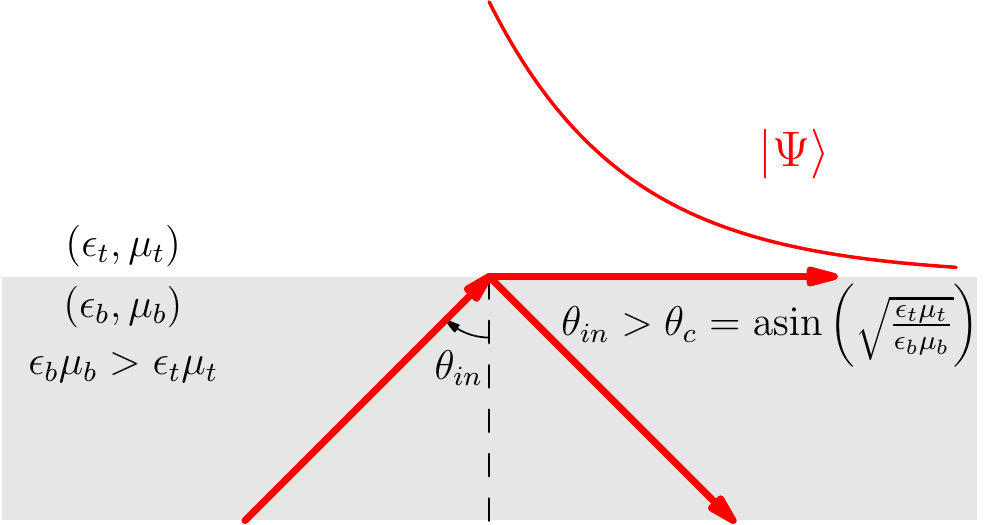}
	\caption{\label{fig:system_nosphere} We consider a planar interface between two media. The permittivity and permeability of the bottom and top media are $(\epsilon_b,\mu_b)$ and $(\epsilon_t,\mu_t)$, respectively. A plane wave propagating in the bottom medium impinges on the interface at an angle $\thi$ larger than the total internal reflection angle $\theta_c$. This creates an evanescent plane wave $|\Psi\rangle$ in the top medium.}
\end{figure}

We now follow this program for the evanescent plane wave depicted in Fig. \ref{fig:system_nosphere}. 

\subsection{Transverse angular momentum content of an evanescent plane wave}
The field $|\Psi\rangle$ on top of the surface in Fig. \ref{fig:system_nosphere} is created by a monochromatic plane wave of frequency $\omega$ incident from below at an angle $\thi$ higher than the angle of total internal reflection $\theta_c$. The field $|\Psi\rangle$ has all the relevant properties of the fields in Fig. \ref{fig:intuition}. Its linear momentum along the $\xhat$ direction is well-defined\footnote{A field with a ``well-defined'' or ``sharp'' component of momentum is a field which is an eigenstate of the corresponding operator. For example, the evanescent plane wave in Fig. \ref{fig:system_nosphere} is an eigenstate of $P_x$, $P_y$, and $P_z$: $P_x|\Psi\rangle=\left(\omega\sqrt{\epsilon_t\mu_t}\right)|\Psi\rangle$, $P_y|\Psi\rangle=0|\Psi\rangle$, $P_z|\Psi\rangle=i\sqrt{p_x^2-\omega^2\epsilon_t\mu_t}|\Psi\rangle$. The total momentum of the field would then be the power of the field times the eigenvalue. In our formulation, the power of a field $|\Phi\rangle$ is proportional to its norm squared $\langle\Phi|\Phi\rangle$. The same considerations apply to other properties like angular momentum.}, and so are the other two momentum components 
\begin{equation}
	\label{eq:p}
	\pp=\left[p_x,0,p_z\right],
\end{equation}
with $p_x^2+p_z^2=\left(\omega\sqrt{\epsilon_t\mu_t}\right)^2$. Additionally, $p_x^2$ is larger than the square of the wavenumber in the top medium 
\begin{equation}
	\label{eq:pxgtk}
	p_x^2=\left(\omega\sqrt{\epsilon_b\mu_b}\right)^2\sin\theta_{in}>\left(\omega\sqrt{\epsilon_t\mu_t}\right)^2,
\end{equation}
where $(\epsilon_b,\mu_b)$ and $(\epsilon_t,\mu_t)$ are the bottom and top medium's permittivity and permeability\footnote{We use units where the vacuum's $\epsilon_0$ and $\mu_0$ are both equal to 1.}, respectively. We assume isotropic and non-absorbing media. As a consequence of Eq. (\ref{eq:pxgtk}), $p_z$ in Eq. (\ref{eq:p}) is imaginary in the top medium,
\begin{equation}
p_z=i\sqrt{p_x^2-\omega^2\epsilon_t\mu_t},
\end{equation}
which confers the field its evanescent character in the $\zhat$ direction: An exponential decrease of the intensity as we go further up the top medium.

Our aim is to decompose the evanescent field $\evfield$ in the top medium into modes of well-defined angular momentum component around the $\yhat$ direction ($J_y$). Additionally, we desire for the angular momentum to be referred to an origin of coordinates located at some distance above the interface between the two media. In this way, the center of coordinates can be made to coincide with the spatial position of a hypothetical particle immersed into the evanescent field, like the sphere in Fig. \ref{fig:system}. Then, the multipolar response of the particle can be used together with the decomposition of the evanescent field to compute the light-matter interaction.

\begin{figure}[h!]
	\includegraphics[width=\linewidth]{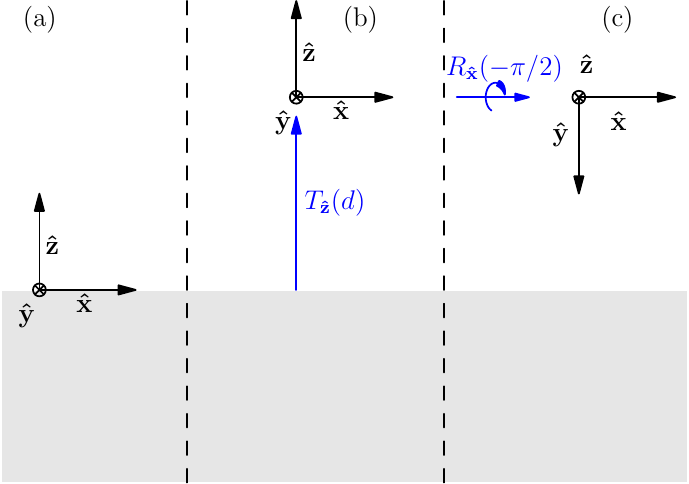}
	\caption{\label{fig:transformations} The figure illustrates the steps that allow to go from a decomposition into multipoles of well-defined $\zhat$ component of angular momentum with origin at the interface (a), to a decomposition into multipoles of well-defined $\yhat$ component of angular momentum with origin at a distance $d$ above the interface (c). We first shift the origin up by a translation along the $\zhat$ axis $T_{\zhat}(d)$ (b), and then (c) rotate the shifted coordinate system by $-\pi/2$ around the $\xhat$ axis $R_{\xhat}(-\pi/2)$.}
\end{figure}

Appendix \ref{sec:jymodes} contains the detailed derivations of the decomposition of $|\Psi\rangle$ into modes with well-defined transverse angular momentum $J_y$. The procedure that we follow in App. \ref{sec:jymodes} is indicated in Fig. \ref{fig:transformations}. First, we obtain the expansion of the evanescent plane wave in a basis of multipolar fields of well-defined $J_z$ with origin at the interface [Fig. \ref{fig:transformations}(a)]. Such a decomposition can be found in \cite{Chew1979,Liu1995,Zvyagin1998}. The key step is to construct the evanescent plane wave by transforming a propagating plane wave with momentum $[0,0,\omega\sqrt{\epsilon_t\mu_t}]$ through a rotation around the $\yhat$ axis by the complex angle \cite{Zvyagin1998,Bekshaev2013}   
\begin{equation}
	\label{eq:ev}
	\theta_{ev} = \arcsin\left(\sqrt{\frac{\epsilon_b\mu_b}{\epsilon_t\mu_t}}\sin\theta_{in}\right)=\arccos \left(\sqrt{1-\frac{p_x^2}{\omega^2\epsilon_t\mu_t}}\right), 
\end{equation}
where the angle of incidence in the bottom medium $\theta_{in}$ is larger than the angle of total internal reflection $\theta_c=\arcsin\left(\sqrt{\frac{\epsilon_t\mu_t}{\epsilon_b\mu_b}}\right)$.

After obtaining the decomposition into modes of well-defined $J_z$, we shift the origin by a distance $d$ above the interface [Fig. \ref{fig:transformations}(b)]. In the last step we rotate the coordinate system to obtain modes of well-defined $J_y$ at the shifted origin [Fig. \ref{fig:transformations}(c)]. 

The final result in App. \ref{sec:jymodes} are the values of the complex amplitudes $\alpha_{j,m_y}^{\pm}$ in the expansion:
\begin{equation}
	\label{eq:jy}
	\evfield=\sum_{\substack{j>0,\\|m_y|\le j}} \alpha_{j,m_y}^+\helplusmy+\alpha_{j,m_y}^-\helminusmy,
\end{equation}
where $|k\ j\ m_y\ \pm\rangle$ denote the multipolar fields of wavenumber $k=\omega\sqrt{\epsilon_t\mu_t}$, angular momentum squared $J^2$ equal to $j(j+1)$ ($j>0$ and integer), angular momentum around the $\yhat$ axis ($J_y$) equal to $m_y$ ($|m_y|\le j$ and integer), and helicity $\lambda=\pm 1$. Appendix A includes instructions for using the publicly available EasySpin code \cite{Stoll2006} to compute $\alpha_{j,m_y}^{\pm}$.

Later in the article we will study the torque induced by evanescent plane waves on spherical absorbing particles. The symmetry of such problem calls for the use of multipolar fields of well-defined parity. 

\begin{equation}
	\label{eq:pmp}
	\evfield=\sum_{\substack{j>0,\\|m_y|\le j}} \beta_{j,m_y}^{\tau=1}|k\ j\ m_y\ \tau=1\rangle +\beta_{j,m_y}^{\tau=-1}|k\ j\ m_y\ \tau=-1\rangle,
\end{equation}

where the parity of $|k\ j\ m_y\ \tau\rangle$ is $\tau(-1)^j$ \cite[Eq. 11.4-7]{Tung1985}. The value $\tau=1$ corresponds to the ``electric'' multipoles, and $\tau=-1$ to the ``magnetic'' multipoles (\cite[Eq. 11.4-25]{Tung1985}, \cite[p. 18]{Berestetskii1982}). 

The relationship between multipoles of well-defined helicity and those of well-defined parity is \cite[Eq. 11.4-6]{Tung1985}:
\begin{equation}
	\label{eq:hpt}
	\sqrt{2}|k\ j\ m_y\ \tau\rangle = |k\ j\ m_y\ +\rangle+\tau|k\ j\ m_y\ -\rangle.
\end{equation}

Given the coefficients of the expansion of $\evfield$ into multipoles of well-defined helicity of Eq. (\ref{eq:jy}), the coefficients for the expansion into multipoles of well-defined parity in Eq. (\ref{eq:pmp}) can be obtained using Eq. (\ref{eq:hpt}):

\begin{equation}
	\label{eq:betaalpha}
	\sqrt{2}\beta_{j,m_y}^{\tau}=\alpha_{j,m_y}^++\tau\alpha_{j,m_y}^-.
\end{equation}

\begin{center}
\begin{figure*}		\subfloat[]{\includegraphics[width=0.5\textwidth]{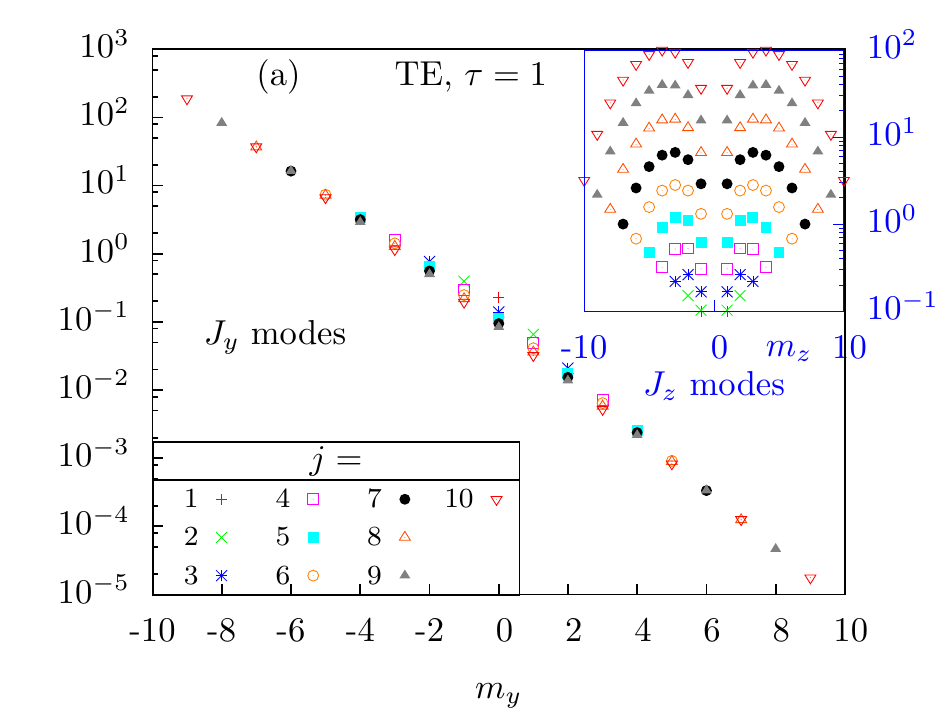}}		\subfloat[]{\includegraphics[width=0.5\textwidth]{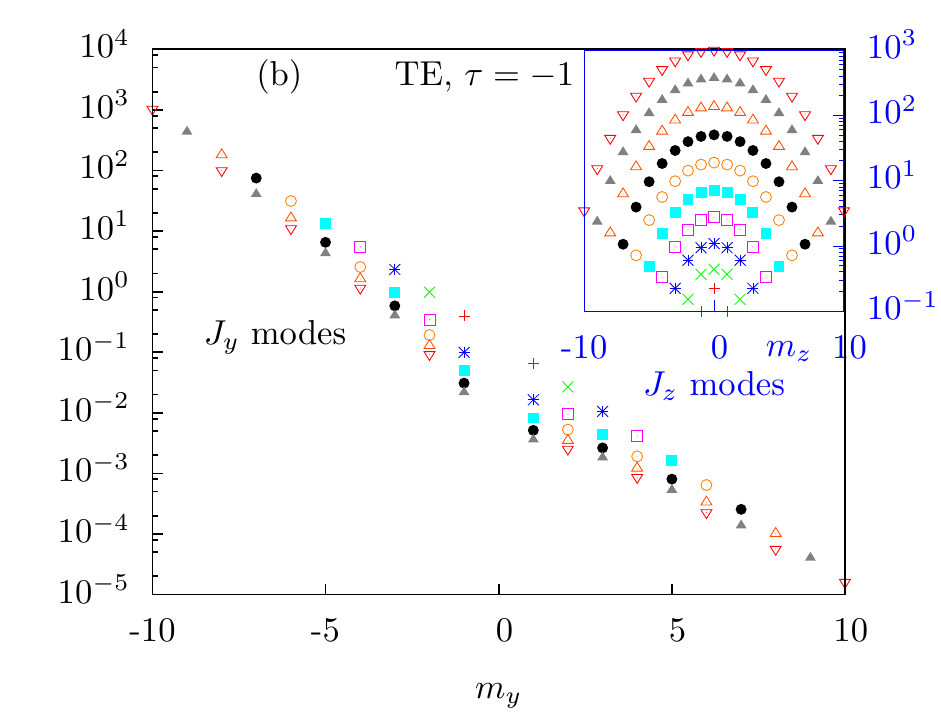}}		\vspace{-0.8cm}
		\subfloat[]{\includegraphics[width=0.5\textwidth]{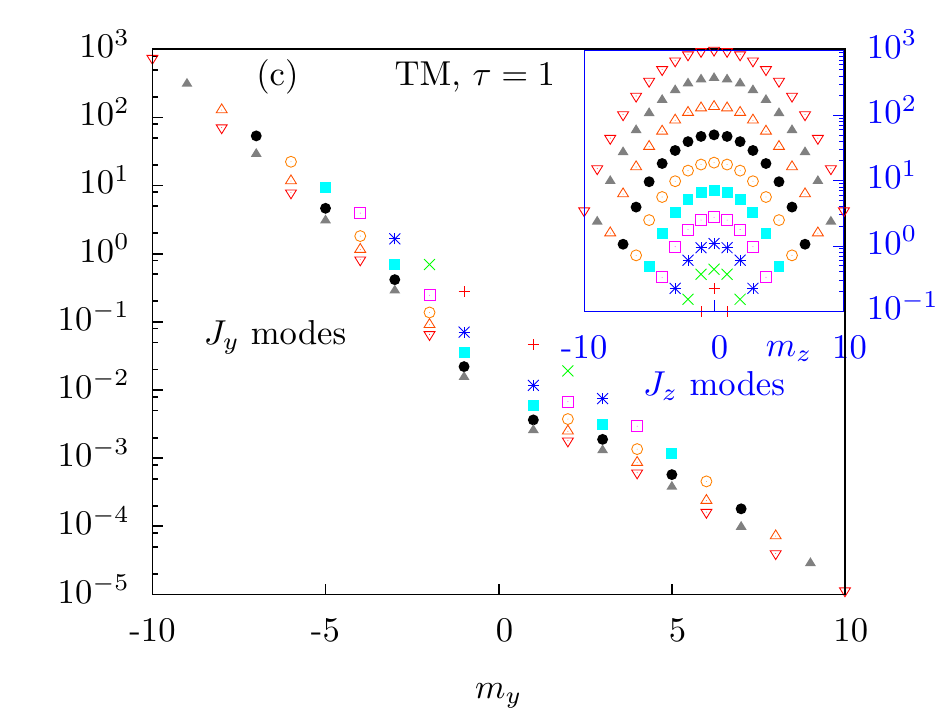}}		\subfloat[]{\includegraphics[width=0.5\textwidth]{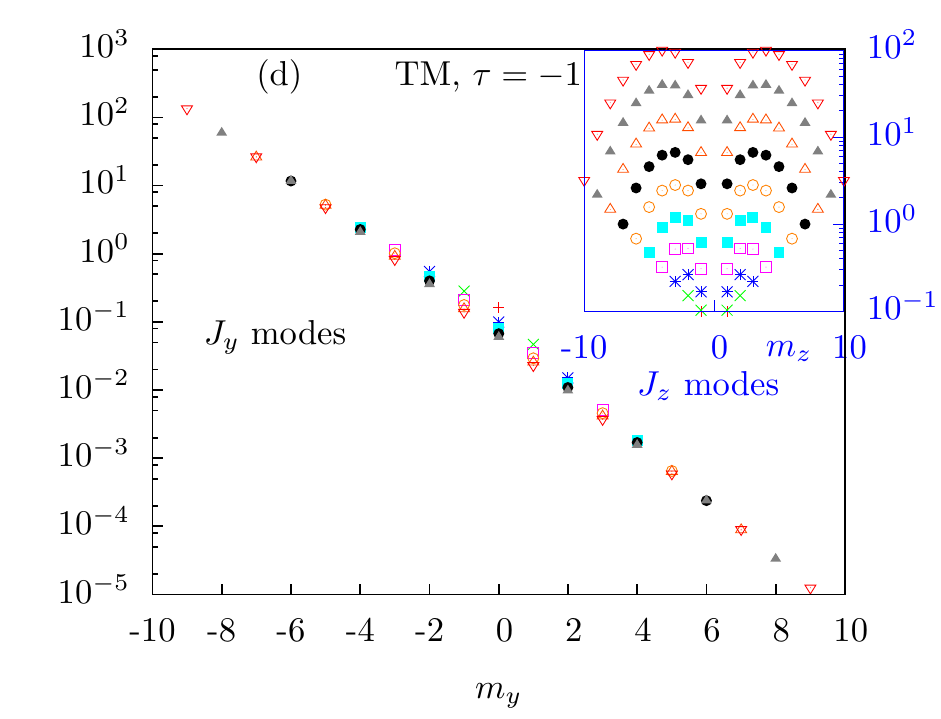}}
	\caption{\label{fig:jyvsjz} Absolute value of the multipolar coefficients of evanescent plane waves in the top medium of Fig. \ref{fig:system_nosphere}. The vertical scale is logarithmic. The key in Fig. \ref{fig:jyvsjz}(a) applies to all plots. The parameters in Fig. \ref{fig:system_nosphere} are set to: $\epsilon_t=\mu_t=\mu_b=1$, $\epsilon_b=2.25$, $\theta_{inc}=1.27$ radians, $d=0.14(2\pi)/(\omega\sqrt{\epsilon_t\mu_t})$. Graphs (a) and (b) correspond to a TE polarized plane wave incident from the bottom medium. Graphs (c) and (d) correspond to a TM polarized plane wave. The four main figures show the amplitudes of a decomposition into multipolar modes with well-defined $J_y$ [$|\beta_{j,m_y}^\tau|$ in Eq. (\ref{eq:pmp})]. The four insets show the amplitudes for multipolar modes with well-defined $J_z$.}
\end{figure*}
\end{center}

The four main graphs in Fig. \ref{fig:jyvsjz} show the amplitudes of the parity coefficients of the decomposition of a TE [Figs. \ref{fig:jyvsjz}(a)-(b)] and a TM [Figs. \ref{fig:jyvsjz}(c)-(d)] evanescent plane wave into modes with well-defined $J_y$. We plot $|\beta_{j,m_y}^{\tau}|$ on a logarithmic scale. The evanescent plane waves result from the setup of Fig. \ref{fig:system_nosphere} with the parameters indicated in the caption of Fig. \ref{fig:jyvsjz}. The infinite range of $j$ in Eq. (\ref{eq:pmp}) is truncated at $j_{max}=10$.

The amplitudes of modes with well-defined $J_y$ show the same trend in the four cases: They grow as $m_y$ decreases. The growth is linear with $-m_y$ on a logarithmic scale, hence exponential on a linear scale. The sign of the slope in the figures is locked to the sign of the $x$ component of linear momentum in Eq. (\ref{eq:p}). Starting with the opposite sign $(p_x\rightarrow-p_x)$ [Fig. \ref{fig:intuition}(b)] results in the opposite slope in all four cases, where modes with large positive $m_y$ dominate. The slope also changes sign when changing the decay direction $\zhat\rightarrow-\zhat$ [Fig. \ref{fig:intuition}(c)]. The simultaneous change of momentum and gradient directions results in the original slope [Fig. \ref{fig:intuition}(d)]. We highlight that these trends are polarization independent: Since they are the same for the TE and the TM plane waves, they will occur in any linear combination of the two. 

The trends in the amplitudes match the intuitive ideas about the situation in Fig. \ref{fig:intuition}. The exponential predominance of amplitudes with a given sign of $m_y$ indicates a predisposition of the field for inducing a torque of the corresponding sense in particles immersed in it. For a given sign of $m_y$, the sense of the torque can be deduced using the right-hand rule.

The match between our intuition and the analytical results is lost in a decomposition into multipoles of well-defined $J_z$. The insets in Fig. \ref{fig:jyvsjz} show the amplitudes of the decompositions into modes with well-defined $J_z$ ($|\beta_{j,m_z}^{\tau}|$) on a logarithmic scale. The amplitudes of the modes with well-defined $J_z$ in the insets are symmetric around $m_z=0$, which does not allow an intuitive prediction of the existence of torque. This shows the advantage of choosing $\yhat$ as the quantization axis for angular momentum in our study. 

Besides matching our intuition, the decomposition into modes of well-defined $J_y$ provides information that can be used in the study and engineering of light-matter interaction for objects located near surfaces. We now discuss the enhanced role that high multipolar orders play in evanescent fields, and a polarization controlled selection rule which exists in mirror symmetric fields like TE and TM plane waves.

\subsection{The enhanced role of high multipolar orders}
\begin{center}
	\begin{figure}
		\includegraphics[width=\linewidth]{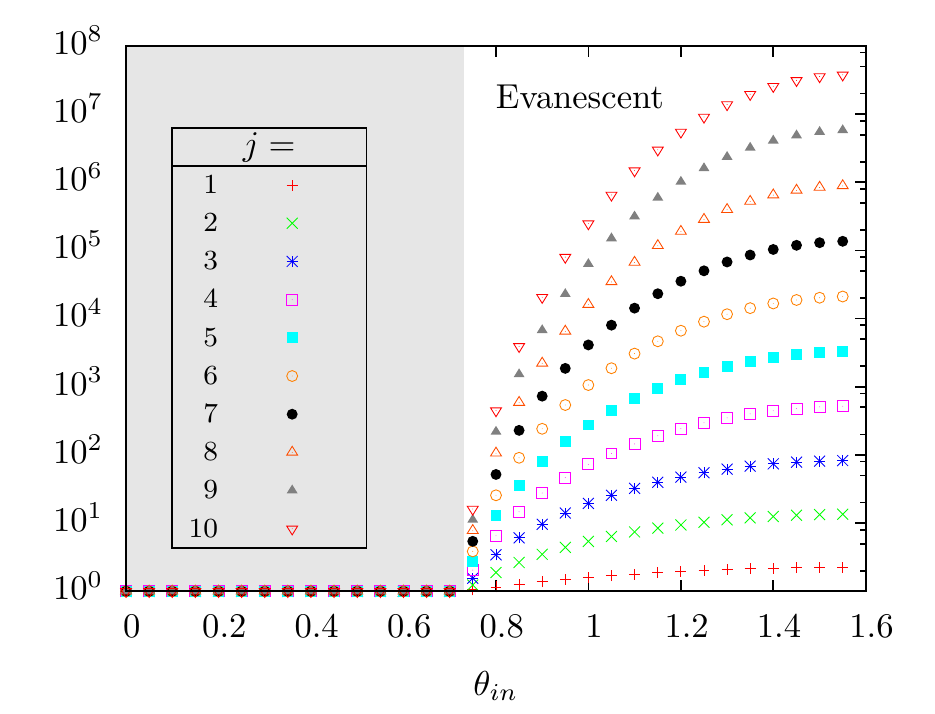}
\caption{\label{fig:djplot} Power (logarithmic scale) of the plane wave in the top medium of Fig. \ref{fig:system_nosphere} contained in the $j$-th multipolar subspace as a function of the angle of incidence $\theta_{in}$ of the plane wave in the bottom medium. The behavior in the propagating region is different than in the evanescent region. In a propagating plane wave, the power in each multipolar space is constant. In an evanescent plane wave, the power concentrates in multipolar subspaces of high $j$. See App. \ref{sec:mj} for the details of the calculations.}
\end{figure}
\end{center}

The modes with large multipolar order $j$ dominate in both $J_y$ and $J_z$ decompositions. This is a distinct characteristic of evanescent plane waves. In the multipolar decomposition of propagating plane waves, all the $j$ subspaces have exactly the same total sum of squared amplitudes. As soon as $\theta_{in}$ is larger than the angle of total internal reflection, the power is unevenly split between the different $j$ subspaces, and it concentrates exponentially at larger $j$ values. This can be appreciated in Fig. \ref{fig:djplot}.
As shown in App. \ref{sec:mj}, the formal origin of this difference is that the angle $\theta_{ev}$ in Eq. (\ref{eq:ev}) is real for propagating plane waves but complex for evanescent plane waves. Notably, this shows that evanescent fields provide a means for enhancing the relative importance of higher multipolar orders in light-matter interactions at close range. This enhancement has been experimentally seen in \cite{Tojo2004}, where the authors could measure atomic quadrupolar transitions of cesium atoms immersed in an evanescent field. The enhancement of the quadrupolar interaction was also theoretically studied in \cite{Tojo2005} by some of the same authors.

\subsection{A polarization controlled selection rule}
In App. \ref{sec:AppC} we show that, given a $(j,m_y)$ pair, the TE plane waves have zero content of one of the two possible parities and the TM plane waves have zero content of the other one. In the transverse multipolar expansion of a TE plane wave, the complex amplitude of the ``electric'' mode $|k\ j\ m_y\ \tau=1\rangle$ contains a multiplicative factor equal to $\left(1-(-1)^{j+m_y}\right)$, and the complex amplitude of the ``magnetic'' mode $|k\ j\ m_y\ \tau=-1\rangle$ a multiplicative factor equal to $\left(1+(-1)^{j+m_y}\right)$. In a TM plane wave, the same multiplicative factors are swapped between the two parities. While the dominance of modes with large transverse angular momentum seen in Fig. \ref{fig:jyvsjz} is polarization independent, the parity content of the evanescent plane wave in each $(j,m_y)$ subspace is not. Table \ref{tab:tetm} shows which polarization produces a non-zero contribution for each value of parity and $m_y$, for the dipolar ($j=1$) and quadrupolar ($j=2$) cases. The consequences of this selection rule can be observed in Fig. \ref{fig:jyvsjz}. 

\begin{table}[h!]
	\begin{ruledtabular}
\begin{center}\begin{tabular}{cccc|ccccc}
	& & {$j=1$} & & & &{$j=2$}  &\\
	\hline
	$m_y$& {-1} & {0} & {1}&{-2} & {-1} & {0}&{1} & {2}\\
	$(\text{electric})\ \tau=1$& {TM} & {TE} & {TM}&{TM} & {TE} & {TM}&{TE} & {TM}\\
	$(\text{magnetic})\ \tau=-1 $& {TE} & {TM} & {TE}&{TE} & {TM} & {TE}&{TM} & {TE}\\
\end{tabular}\end{center}
	\end{ruledtabular}
	\caption{\label{tab:tetm}Polarization of the plane wave with non-zero contribution in the $(j,m_y)$ multipolar mode of parity $\tau(-1)^j$. The table is valid for any field that is an eigenstate of the mirror reflection across the XZ plane. The behavior of fields with eigenvalue -1(+1) is given by that of the TE(TM) plane wave.}
\end{table}

Crucially, the only relevant assumption in the derivation of the selection rule is the fact that the TE(TM) plane waves are eigenstates of the mirror reflection across the XZ plane with opposite eigenvalue -1(+1). As argued at the end of App. \ref{sec:AppC}, the rule applies not only to plane waves, but also to general eigenstates of this particular mirror reflection. The rule is hence relevant for the selective excitation of resonances in objects placed in general evanescent fields with mirror symmetry, like for example the TE and TM modes of an optical fiber. 

Incidentally, Fig.  \ref{fig:jyvsjz} shows another regularity: The coefficients for modes with $m_y=0$ are equal to zero for all even values of $j$. This can be traced back to a regularity of the small Wigner matrices, and is actually a feature of all fields with zero eigenvalue of the $\yhat$ component of linear momentum.

\section{Application to experiments\label{sec:exp}}
The analysis and results contained in Sec. \ref{sec:tjy} can be applied to the analysis of evanescent light-matter interactions. We now discuss two such applications.

\subsection{Directional coupling of emitters into evanescent modes\label{sec:direc}}
The experimental results reported in Refs. \cite{Petersen2014,Rodriguez2014,Sollner2015} provide very clear evidence of evanescent directional coupling of an emitter into guided modes. The origin and magnitude of this directionality can be elucidated using the analysis from Sec. \ref{sec:tjy}.

Let us consider an electric dipolar emitter located above the interface and exclusively generating {\protect $|k\ \ j=\nolinebreak 1\ \ m_y=\nolinebreak-1\ \ \tau=\nolinebreak1\rangle$} waves. We are first interested in the coupling of the dipole emission into the plane waves analyzed in Fig. \ref{fig:jyvsjz}, and into their counter-propagating versions with $p_x<0$. The coupling efficiency of the dipole into each plane wave will be proportional to the amplitude squared of the coefficient $\beta_{1,-1}^{\tau=1}$ in the corresponding expansion. This statement follows from electromagnetic reciprocity.

Let us first apply the parity selection rule in Tab. \ref{tab:tetm}: The electric dipole $(j=1,m_y=-1)$ will only couple to a TM mode. Then, for the considered TM plane waves, the ratio between $|\beta_{1,-1}^{\tau=1}|^2$ for $p_x>0$ and  $p_x<0$, and hence the ratio between the coupling efficiency to the two counter-propagating TM plane waves is 
\begin{equation}
	\label{eq:j1m1}
	\frac{\left|\beta_{j=1,m_y=-1}^{\tau=1,p_x>0}\right|^2}{\left|\beta_{j=1,m_y=-1}^{\tau=1,p_x<0}\right|^2}=\frac{\left|\beta_{j=1,m_y=-1}^{\tau=1,p_x>0}\right|^2}{\left|\beta_{j=1,m_y=1}^{\tau=1,p_x>0}\right|^2}\approx 36,
\end{equation}
where the first equality follows from equality of the denominators due to symmetry.

Equation (\ref{eq:j1m1}) means that the electric $(j=1,m_y=-1)$ dipole will couple to the plane wave with $p_x>0$ with 36 times more strength than to the one with $p_x<0$. The directionality is exactly reversed for the $(j=1,m_y=1)$ dipole. Overall, the sign of the transverse angular momentum of the dipole emission determines the directionality of the excited wave to a large extent. Notably, the directionality ratio increases exponentially with the multipolar order. It is equal to 36 for the $(j=1,m_y=\pm 1)$ modes, equal to 1340 for $(j=2,m_y=\pm 2)$, and equal to 5$e$4 for $(j=3,m_y=\pm 3)$. Similar analysis and conclusions would apply to the directional coupling of a magnetic dipole into TE modes (see Tab. \ref{tab:tetm}).

The above analysis uses both the exponential behavior of the coupling coefficients and the parity selection rule for mirror symmetric fields. It applies directly to Refs. \cite{Petersen2014,Rodriguez2014}, where the guided modes have the appropriate mirror symmetry. In Ref. \cite{Sollner2015}, the waveguide breaks the symmetry and the parity selection rule cannot be used. Nevertheless, the polarization independent exponential dependence of the coupling coefficients creating the directionality still applies. 

The rigorous quantitative analysis of the experiments in Refs. \cite{Petersen2014,Rodriguez2014} can be achieved by considering the decomposition of the actual optical guided mode, instead of the plane waves in Fig. \ref{fig:jyvsjz}.

We note that the directionality is achieved by means of angular momentum selection, not handedness selection. When an emitter produces a pure helicity wave of the kind $|k\ j\ m_y\ \lambda\rangle$, its preferential coupling direction is completely determined by the transverse angular momentum $m_y$, and is independent of the helicity $\lambda$. We can illustrate this with the following example: A $|k\ \ j=1\ \ m_y=-1\ \ \lambda\rangle$ wave preferentially couples towards one direction. If we rotate the emitter by a rotation that takes $\yhat$ to $-\yhat$, the handedness of the radiation does not change, but its angular momentum flips sign, and it will now preferentially couple towards the opposite direction.

This is an important distinction. An electric or magnetic dipolar emitter with $m_y=\pm1$ can be associated with the vectors $(\mp\zhat- i\xhat)/\sqrt{2}$, which suggests the name ``circularly polarized'' electric or magnetic dipoles. This does not mean that they emit circularly polarized waves of a single handedness, which is what dipoles of well-defined helicity do \cite[Sec. 2.7.3]{FerCorTHESIS}, \cite{FerCor2013,Zambrana2016Nano}. Dipoles of well-defined helicity require the simultaneous presence of electric and magnetic dipoles in a definite phase relationship, as can be deduced from inverting Eq. (\ref{eq:betaalpha}). 

\subsection{Excitation of atoms into opposite Zeeman states}
We finish the section with a discussion of the simultaneous excitation of atoms into opposite Zeeman states by means of the same guided optical mode \cite{Mitsch2014}. The experiments in Ref. \cite{Mitsch2014} show that the atoms trapped on ``top'' of a tapered fiber preferentially make a dipolar transition of opposite angular momentum with respect to those trapped at the ``bottom'' of the fiber. 

We note that the intensity of the evanescent fields that interact with the two groups of trapped atoms decreases in opposite directions (say $\zhat$ and $-\zhat$). As can be appreciated in Fig. \ref{fig:intuition}, this gradient reversal has the same effect as the change in propagation direction: It reverses the slopes in Fig. \ref{fig:jyvsjz}. This means that the sign of the transverse angular momentum of the dominant modes will be opposite in the two cases, and that the electric dipole transition of atoms on ``top'' will preferentially be of the opposite sign of transverse angular momentum than for the atoms at the ``bottom''. This transition rate difference is of the same character as the ratio in Eq. (\ref{eq:j1m1}). If we consider the reflection across the plane containing both the ``top'' and ``bottom'' atoms, both the TM plane wave used in Eq. (\ref{eq:j1m1}) and the illumination injected in the tapered fiber have\footnote{This can be deduced from \cite[Fig. 2]{Reitz2014}.} eigenvalue 1. Using the coefficients of the transverse multipolar expansion of the actual optical mode in the fiber will result in the rigorous prediction of the transition rate difference.

The outcomes of our symmetry based arguments are consistent with those of the microscopic theory for the forward and backward scattering of guided light from a multilevel atom developed in \cite{Le2014}.

Notably, the rate difference will increase exponentially for higher order multipolar transitions. This is essentially due to the same reason that produces the directionality explained in Sec. \ref{sec:direc}: The exponential trend shown by the decomposition of evanescent plane waves with respect to the transverse angular momentum (e.g. Fig. \ref{fig:jyvsjz}).

\section{Transverse torque on spherical particles in evanescent fields\label{sec:torque}}

\begin{figure}[h!]
	\includegraphics[width=\linewidth]{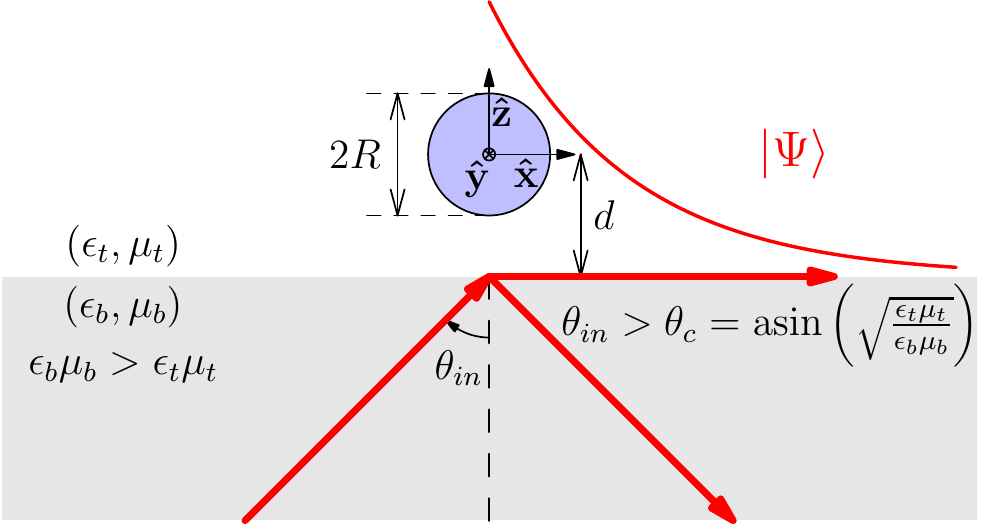}
	\caption{\label{fig:system} An absorbing spherical particle of radius $R$ located at a distance $d$ above the interface interacts with the evanescent field produced in the system of Fig. \ref{fig:system_nosphere}. }
\end{figure}

In this section, we will use the decomposition of Sec. \ref{sec:tjy} to analyze and maximize the transverse torque experienced by a spherical absorbing particle immersed in an evanescent plane wave. The sketch of the problem is displayed in Fig. \ref{fig:system}. A sphere of radius $R$ located at a distance $d$ above the interface interacts with the evanescent field created by the total internal reflection of a plane wave propagating in the bottom medium.

In order to solve this electromagnetic problem exactly, the Green tensor of the bilayer needs to be used to account for the interaction between the particle and the environment. The initial field incident on the particle is the evanescent plane wave. The particle interacts with it and generates a scattered field which reflects back from the surface to the particle. The particle interacts with this new field, and so on. It is however very common \cite{Chew1979,Almaas1995,Bekshaev2013,Almaas2013,Wang2014,Canaguier-Durand2014,Antognozzi2016} to ignore the back action from the particle onto itself, and assume that the particle is effectively placed in an homogeneous medium where the initial evanescent plane wave is the only incident field. As argued in \cite{Chew1979}, the approximation improves as the particle is located further from the interface. Our choice of parameters $d=700$nm and $R=70$nm are more conservative than in \cite{Chew1979}.

Under this approximation, the problem can be solved using Mie theory. The Mie problem can be solved using the multipoles of well-defined helicity as it has been done in \cite{Zambrana2013,Zambrana2016Nano}. Nevertheless, the problem is usually solved in the parity basis, which is adapted to the symmetry of an achiral spherical particle. Thus, we will use the expansion of the evanescent wave into multipoles of well-defined parity of Eq. (\ref{eq:pmp}). In this section we will use the symbols $(e)$ for ``electric'', and $(m)$ for ``magnetic'',  to denote quantities related to $\tau=+1$ and $\tau=-1$ in Eq. (\ref{eq:pmp}), respectively. 

In Mie theory, the scattering problem is solved once the expression of the incident electromagnetic field is decomposed into multipoles:
\begin{equation}
	\Ei = \displaystyle\sum_{\substack{j>0,\\|m_i|\le j}} \ajie \Ae + \ajim \Am,
\end{equation}
where the subindex $i$ will be equal to either $y$ or $z$.

Then, once the coefficients $\ajiy$ are known, the scattered field can be immediately found as \cite{Bohren1983,ZambranaThesis}:
\begin{equation}
	\Es = \displaystyle\sum_{\substack{j>0,\\|m_i|\le j}} \ajie a_j \Ae + \ajim b_j \Am,
\end{equation}
where $(a_j,b_j)$ are the so-called Mie coefficients. The Mie coefficients are independent of the choice of the angular momentum quantization axis. 

\subsection{Expression of the torque in the basis of transverse angular momentum multipoles\label{sec:mietheory}}
The knowledge of the total field $(\Ei+\Es)$ allows us to calculate Maxwell's stress-energy tensor, which then allows us to compute the net time-averaged radiation torque $\mathbf{N}$ on the sphere via the corresponding flux integrals. 

If the Mie problem is solved in the multipolar basis of well-defined $J_z$, the torques on the sphere can be computed using the formulas obtained by Barton \textit{et al.} \cite{Barton1989}, which were later corrected by Farsund \textit{et al.} in \cite{Farsund1996}. The torques are normalized by $R^3|\Phi_{inc}|^2$, yielding $\mathbf{N}$ in dimensionless units:

\begin{widetext}
	{\small
\begin{eqnarray}
\label{eq:Nxmz}
N_{x,m_z}  &=&   \dfrac{-kR}{8 \pi} \displaystyle \sum_{\substack{j>0,\\|m_z|\le j}} \mathrm{Re} \left\lbrace \sqrt{(j-m_z)(j+m_z+1)} \displaystyle\left[ \left( \vert a_j \vert^2 - \mathrm{Re} \left\lbrace a_j  \right\rbrace \right) \beta_{j,m_z}^{(e)} \beta_{j,m_z+1}^{(e) *} + \left( \vert b_j \vert^2 - \mathrm{Re} \left\lbrace b_j  \right\rbrace \right)\beta_{j,m_z}^{(m)} \beta_{j,m_z+1}^{(m) *} \ \right]  \right\rbrace, \\
N_{y,m_z}  &=&   \dfrac{-kR}{8 \pi} \displaystyle \sum_{\substack{j>0,\\|m_z|\le j}} \mathrm{Im} \left\lbrace \sqrt{(j-m_z)(j+m_z+1)} \displaystyle\left[ \left( \vert a_j \vert^2 - \mathrm{Re} \left\lbrace a_j  \right\rbrace \right) \beta_{j,m_z}^{(e)} \beta_{j,m_z+1}^{(e) *} + \left( \vert b_j \vert^2 - \mathrm{Re} \left\lbrace b_j  \right\rbrace \right)\beta_{j,m_z}^{(m)} \beta_{j,m_z+1}^{(m) *}  \right]  \right\rbrace, \\
\label{eq:Nzmz}
N_{z,m_z} &=&  \dfrac{-kR}{8 \pi} \displaystyle \sum_{\substack{j>0,\\|m_z|\le j}} m_z \left[ \left( \vert a_j \vert^2 - \mathrm{Re} \left\lbrace a_j  \right\rbrace \right) \vert \beta_{j,m_z}^{(e)} \vert^2  + \left( \vert b_j \vert^2 - \mathrm{Re} \left\lbrace b_j  \right\rbrace \right) \vert \beta_{j,m_z}^{(m)} \vert^2 \right].
\end{eqnarray}
}
\end{widetext}
where it is assumed that $\epsilon_t=\mu_t=1$ for the top medium. Note that Eqs. (\ref{eq:Nxmz}-\ref{eq:Nzmz}) differ from those written in \cite{Barton1989,Farsund1996,Canaguier-Durand2014} by a $j(j+1)$ factor because we always consider multipolar fields of unit norm, like those in \cite{Tung1985,Timo2014}.

In addition, the terms in Eqs. (\ref{eq:Nxmz}-\ref{eq:Nzmz}) have been regrouped in a different way in order to highlight the torque dependence on $\left( \vert a_j \vert^2 - \mathrm{Re} \left\lbrace a_j  \right\rbrace \right) $ and $\left( \vert b_j \vert^2 - \mathrm{Re} \left\lbrace b_j  \right\rbrace \right) $. We define these quantities as the \textit{absorption} Mie coefficients:

\begin{equation}
	\label{eq:abs}
	\begin{split}
		A_j&=  \vert a_j \vert^2 - \mathrm{Re} \left\lbrace a_j  \right\rbrace, \text{ and}\\
		B_j&= \vert b_j \vert^2 - \mathrm{Re} \left\lbrace b_j  \right\rbrace. 
	\end{split}
\end{equation}
This definition is motivated by their appearance in Eqs. (\ref{eq:Nxmz}-\ref{eq:Nzmz}) [and later in Eqs. (\ref{eq:Nxmy}-\ref{eq:Nzmy})], the typical definition of absorption cross section in Mie theory, \textit{i.e.} $Q_{abs}=Q_{ext}-Q_{sca}$ \cite{Bohren1983}, and the following consideration. Since scattering off the sphere preserves the angular momentum of multipolar modes, absorption is the only mechanism that allows the electromagnetic field to exert torque on a spherical particle. The conservation of the angular momentum of the whole system (electromagnetic field-sphere), dictates that when the sphere absorbs radiation with a value of $J_y$ equal to $m_y$, it must experience a torque proportional to $m_y$. Furthermore, the torque must be proportional to the absorption. Here, it is seen that if $A_j=B_j=0 \ \forall j$, the torque exerted on the particle is going to be null. This is exactly the case when the permittivity and permeability of the sphere are real, and there are no absorption losses \cite{Bohren1983}. 

In our case, we need the formulas corresponding to Eqs. (\ref{eq:Nxmz}-\ref{eq:Nzmz}), but for multipolar fields of well-defined $J_y$. They can be obtained using the rotation described in Fig. \ref{fig:transformations} (c). No calculations need to be done to find out the analytic expressions of $\mathbf{N}$ in the desired basis. Due to the fact that the scatterer is a sphere, it can be proven that the expressions for $N_x,N_y,$ and $N_z$ as a function of $m_y$ are:
\begin{widetext}
	{\small
\begin{eqnarray}
	\label{eq:Nxmy}
	N_{x,m_y}  &=&   \dfrac{-kR}{8 \pi} \displaystyle \sum_{\substack{j>0,\\|m_y|\le j}} \mathrm{Re} \left\lbrace \sqrt{(j-m_y)(j+m_y+1)} \displaystyle\left[ \left( \vert a_j \vert^2 - \mathrm{Re} \left\lbrace a_j  \right\rbrace \right) \beta_{j,m_y}^{(e)} \beta_{j,m_y+1}^{(e) *} + \left( \vert b_j \vert^2 - \mathrm{Re} \left\lbrace b_j  \right\rbrace \right)\beta_{j,m_y}^{(m)} \beta_{j,m_y+1}^{(m) *} \ \right]  \right\rbrace, \\\label{eq:Nxmyactual}
	N_{y,m_y}  &=&  \dfrac{-kR}{8 \pi} \displaystyle \sum_{\substack{j>0,\\|m_y|\le j}} m_y \left[ \left( \vert a_j \vert^2 - \mathrm{Re} \left\lbrace a_j  \right\rbrace \right) \vert \beta_{j,m_y}^{(e)} \vert^2  + \left( \vert b_j \vert^2 - \mathrm{Re} \left\lbrace b_j  \right\rbrace \right) \vert \beta_{j,m_y}^{(m)}\vert^2 \right], \\
	N_{z,m_y} &=&  \dfrac{kR}{8 \pi} \displaystyle \sum_{\substack{j>0,\\|m_y|\le j}} \mathrm{Im} \left\lbrace \sqrt{(j-m_y)(j+m_y+1)} \displaystyle\left[ \left( \vert a_j \vert^2 - \mathrm{Re} \left\lbrace a_j  \right\rbrace \right) \beta_{j,m_y}^{(e)} \beta_{j,m_y+1}^{(e) *} + \left( \vert b_j \vert^2 - \mathrm{Re} \left\lbrace b_j  \right\rbrace \right)\beta_{j,m_y}^{(m)} \beta_{j,m_y+1}^{(m) *}  \right]  \right\rbrace.
\label{eq:Nzmy}
\end{eqnarray}
}
\end{widetext}

Note that Eqs. (\ref{eq:Nxmz}-\ref{eq:Nzmy}) can be used to verify that the coefficients $\beta_{j,m_z}^{(\tau)}$ and $\beta_{j,m_y}^{(\tau)}$ displayed in Fig. \ref{fig:jyvsjz} are properly related. We can indeed verify that $N_{i,m_z}=N_{i,m_y}$ for $i=x,y,z$.

\subsection{Torque maximization}
The torque around the $\yhat$ axis exerted on a sphere by any plane wave can be calculated using Eq. (\ref{eq:Nxmyactual}) and the coefficients $\beta_{j,m_y}^{(\tau)}$ from Eq. (\ref{eq:pmp}). This analytical formula allows us to study how a given parameter affects the torque. For example, the images in Fig. \ref{fig:intuition} provide some intuition regarding the existence and the sign of a transverse torque on a particle. However, the dependence of the transverse torque $N_y$ on the incidence angle $\thi$ is not clear. 

Equation (\ref{eq:Nxmyactual}) can also be used to find the wavelength, polarization, and angle of incidence of the normalized monochromatic plane wave propagating in the bottom medium of Fig. \ref{fig:system}, which maximizes the torque on a given sphere in the top medium. To such end, we probe a spherical particle using the excitation scheme depicted in Fig. \ref{fig:system} with plane waves of different wavelengths varying from 380nm to 625nm, and different incidence angles $\thi$ varying from $\theta_{in}=\theta_c$ to $\theta_{in}=\pi/2$. Both TE and TM polarizations are considered for the incident plane wave in the bottom medium. All the different plane waves in the bottom medium are assumed to have the same intensity. The optical constants of the top and bottom media are $\epsilon_t=\mu_t=\mu_b=1$, and $\epsilon_b=2.25$. The size of the sphere is $R=70$nm, the distance from the center of the sphere to the interface is $d=700$nm. 

Two different particles are considered, one made of gold and the other made of silicon. The permittivity of gold and silicon are obtained through linear interpolation of the data points in \cite{Palik1998} and \cite{Vuye1993}, respectively. Both gold and silicon absorb in the range of wavelengths that we consider.

\begin{center}
\begin{figure*}
		\subfloat[]{\includegraphics[width=0.5\textwidth]{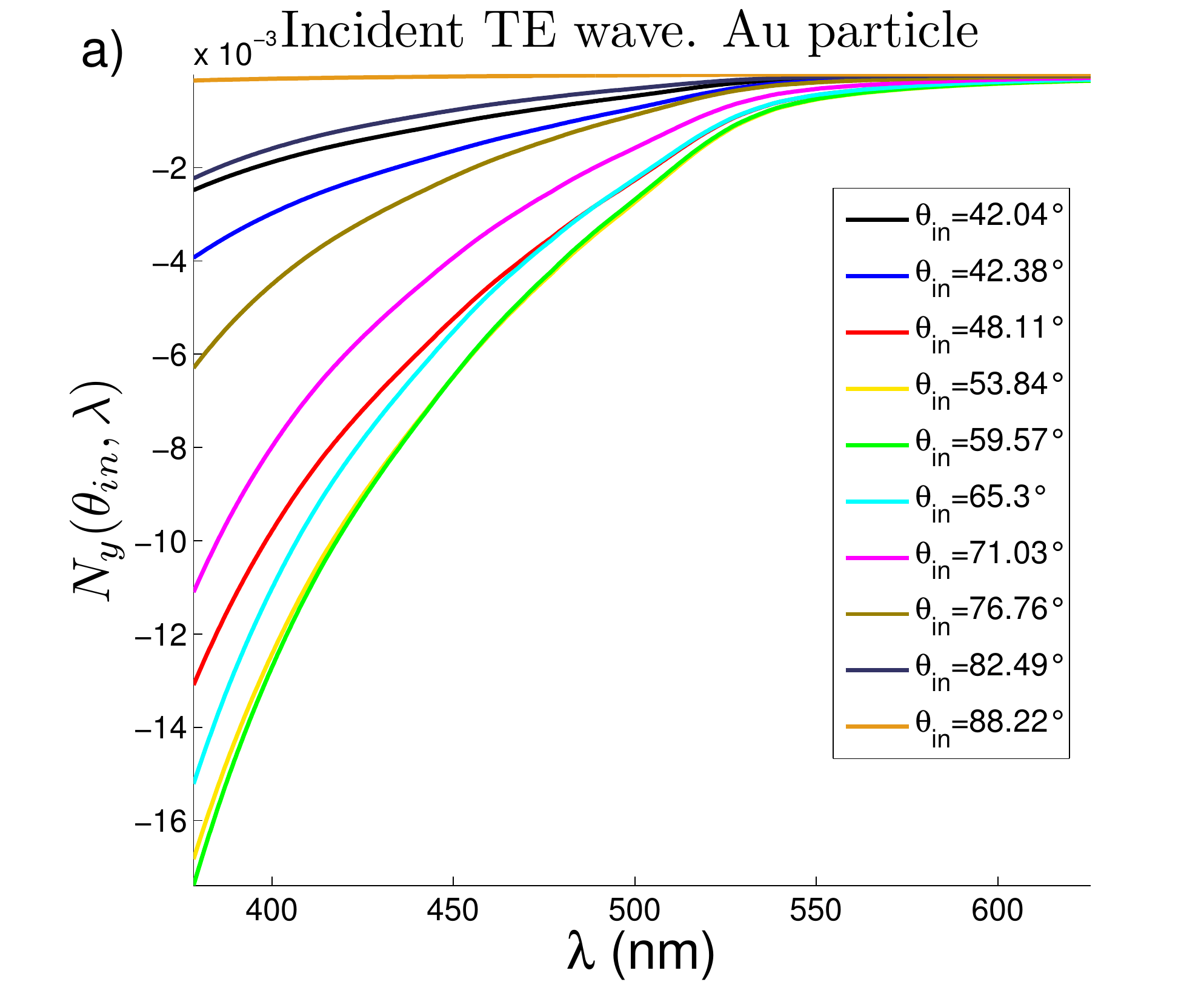}}		\subfloat[]{\includegraphics[width=0.5\textwidth]{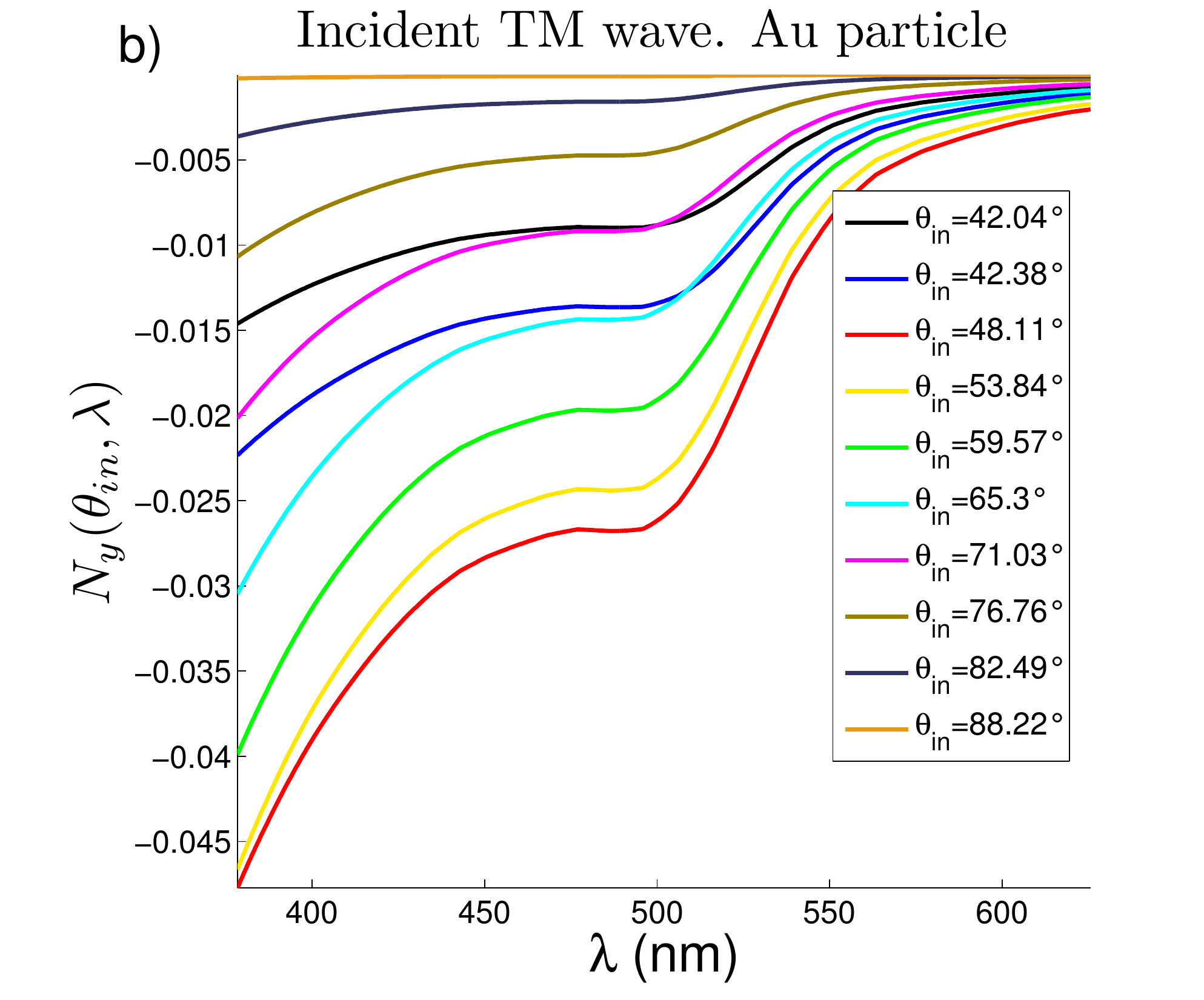}}\\%
		\subfloat[]{\includegraphics[width=0.5\textwidth]{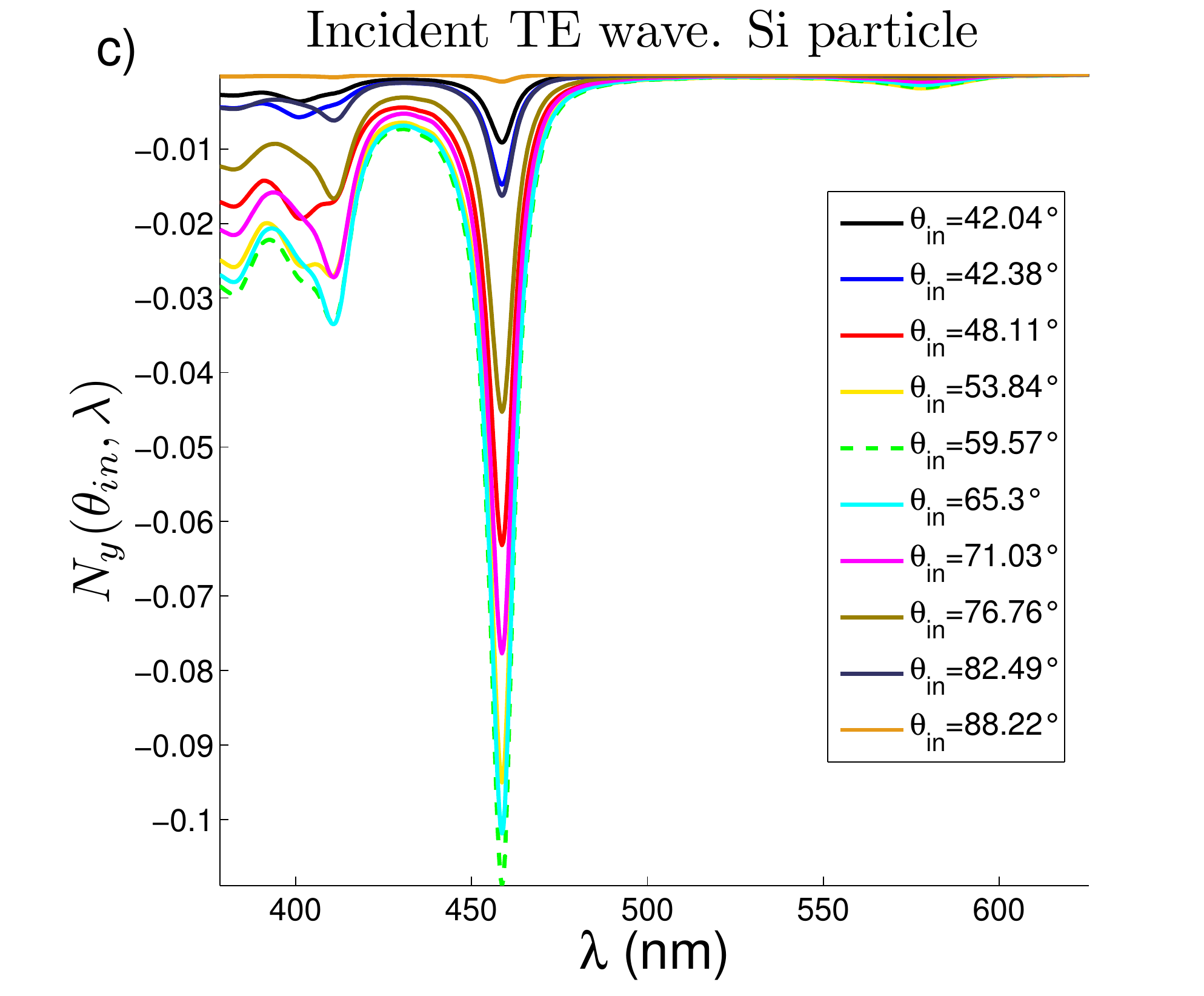}}		\subfloat[]{\includegraphics[width=0.5\textwidth]{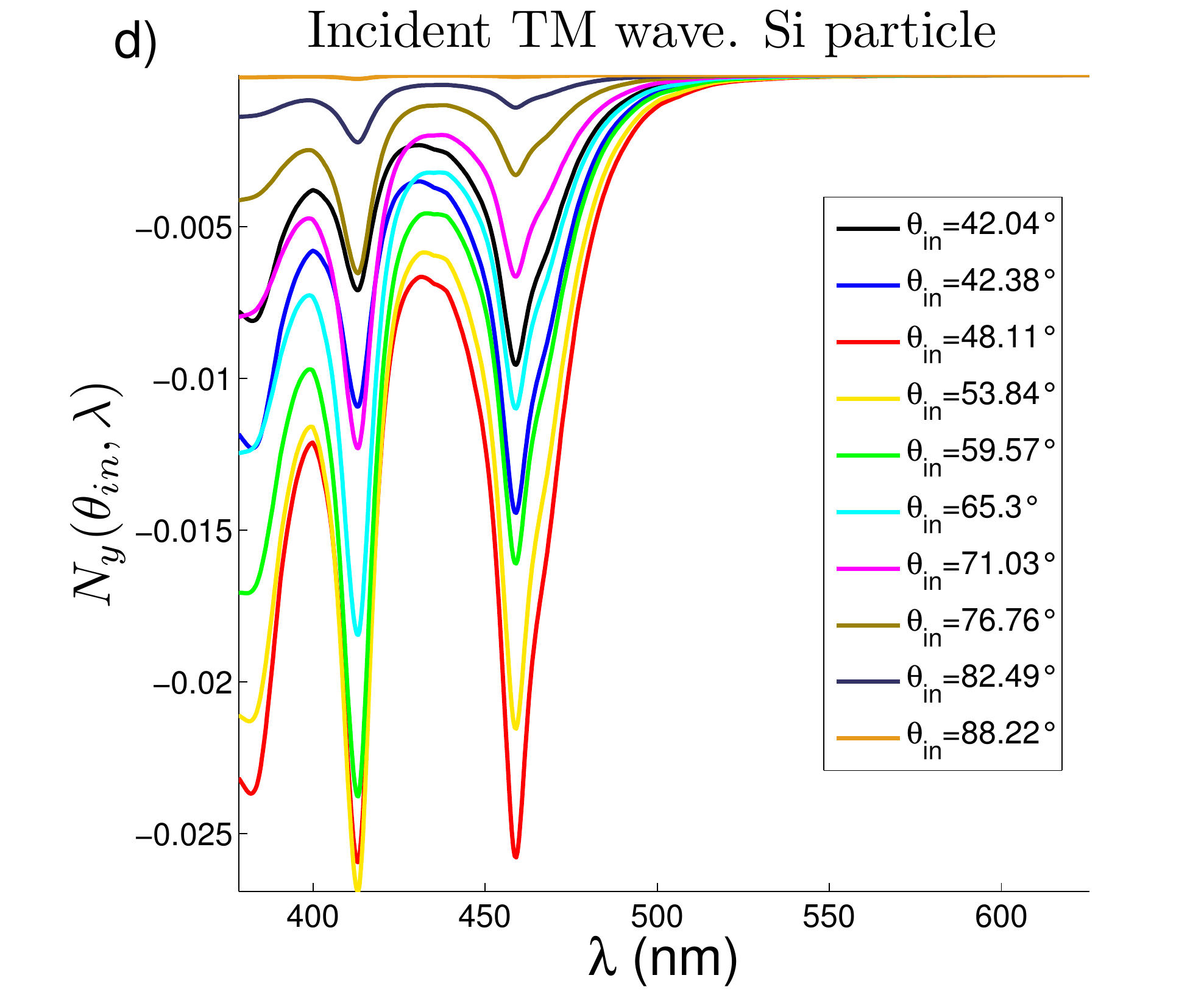}}		\caption{\label{fig:angle} Torque induced on a particle under evanescent plane wave illumination. The geometry of the problem is described by Fig. \ref{fig:system}, with $R=70$nm, $d=700$nm, $\epsilon_t=\mu_t=\mu_b=1$, and $\epsilon_b=2.25$. The plane wave in the bottom medium has a TE polarization for panels (a) and (c), and a TM polarization for panels (b) and (d). The particle is made of gold for panels (a) and (b), and it is made of silicon for panels (c) and (d). For all the plots, different incident angles $\theta_{in}$ have been considered, going from $\theta_c$ to $\pi/2$.  The torques are dimensionless (see text).}
\end{figure*}
\end{center}

Figure \ref{fig:angle} shows the results obtained for the transverse torque $N_y$ for a particle made of gold and (a) an incident TE wave, and (b) an incident TM wave. In Figs. \ref{fig:angle} (c) and (d), the same calculations are done for a particle made of silicon. First, we note that $N_y$ is negative in all cases. This is consistent with the fact that the setting considered in this section corresponds to the situation in Fig. \ref{fig:intuition} (a). We also observe that $N_y$ is maximized for certain angles $\thi$. The optimal angles are found to belong to the $47^{\circ}-66^{\circ}$ interval. Note that these optimal angles are a function of different variables. That is, the optimal angles vary depending on the wavelength, polarization (TE or TM), and material (gold or silicon). This fact implies that the optimum angle is not the result of the difference in transmission given by the $\shat$ and $\phat$ Fresnel coefficients.

We observe that $N_y$ exhibits a clear resonant behavior for the silicon particle. Silicon spheres are known to have a rich resonant behavior \cite{Aitzol2011,Shi2012,Zambrana2015,Muljarov2014}. The interplay between the resonances of the sphere and the selection rule illustrated in Tab. \ref{tab:tetm} will be the focus of the next section.

With respect to maximizing the torque in the Si sphere, we determine from Figs. \ref{fig:angle}(c) and (d) that a TE plane wave with a wavelength around $458$nm and angle of incidence equal to $59.57^{\circ}$ is the monochromatic plane wave propagating in the bottom medium whose evanescent field results in the maximum torque on the silicon particle. We display its torque with a dashed line in Fig. \ref{fig:angle}(c). This is opposite to the case of propagating plane waves, where the optimal polarization for exerting torque on spherical absorbing particles is the circular polarization \cite{Marston1984,Aso2016}, i.e. the equal amplitude mix of TE and TM waves.

\subsection{Selective excitation of absorption resonances}
Looking at Figs. \ref{fig:angle} (c) and (d), one can see that incident $\shat$ and $\phat$ evanescent waves give rise to very different torques. Now, the peaks seem to be linked with resonances of the silicon sphere. Different peaks seem to arise depending on the polarization of the incident evanescent wave. 

\begin{center}
\begin{figure*}
\includegraphics[width=0.9\textwidth]{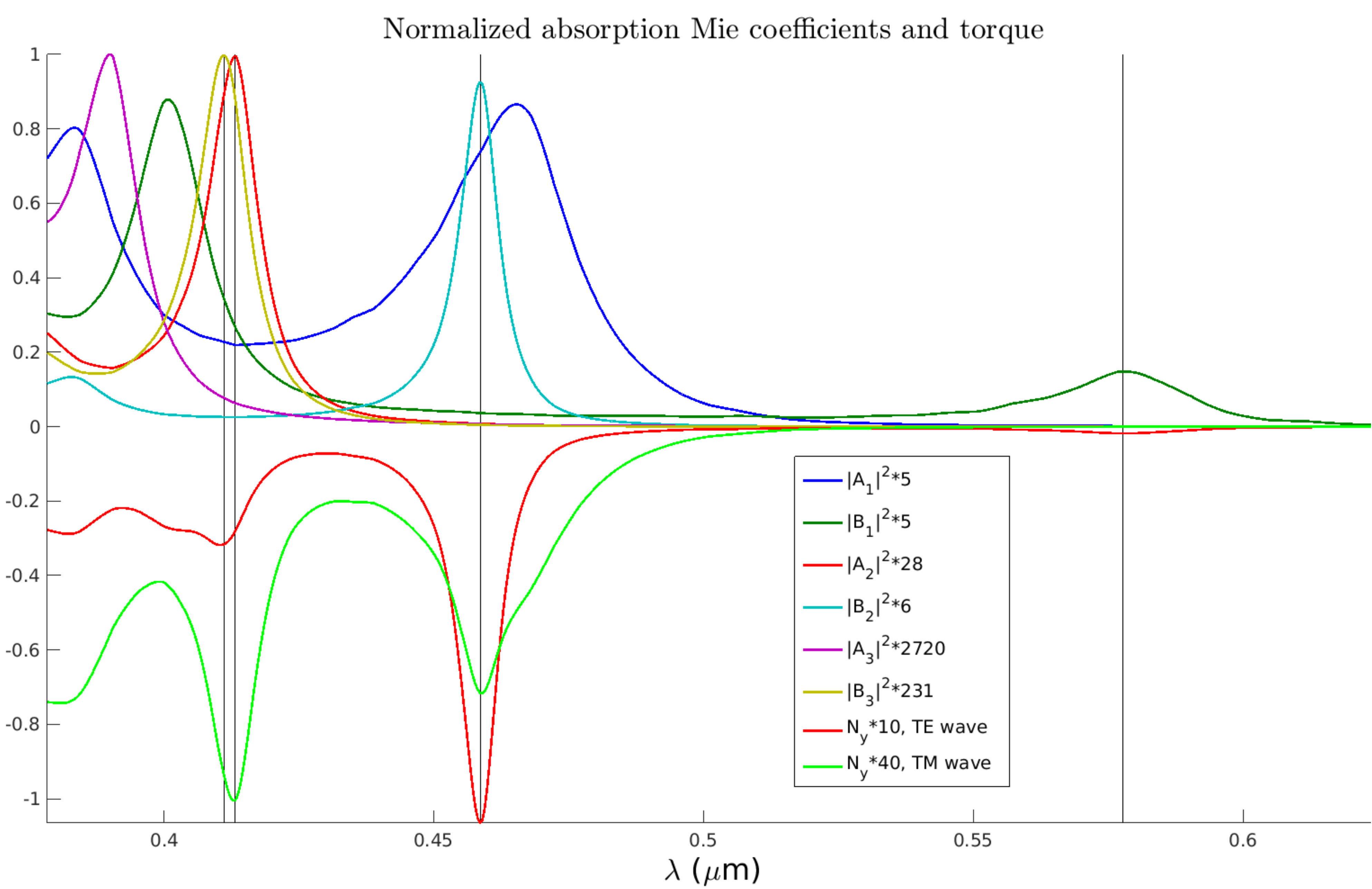}
\caption{\label{fig:sp} Modulus square of the \textit{absorption} Mie coefficients $A_j,B_j$ with $j=1,2,3$ for a $R=70$nm silicon sphere embedded in air. These coefficients are compared with the torque experienced by the same silicon particle when it is located at $d=700$nm above the interface. The value of the torque has been computed for incident TE and TM plane waves with $\thi=57.6^{\circ}$. The torques are dimensionless (see text).}
\end{figure*}
\end{center}

In order to analyze these features, in Fig. \ref{fig:sp} we plot $N_y(\lambda)$ for $\thi=57.6^{\circ}$ for $\shat$ and $\phat$ waves. The angle belongs to the optimal range found in the previous section. Then, we also plot the first three \textit{absorption} Mie coefficients, $A_j,B_j$ of Eq. (\ref{eq:abs}) for $j=1,2,3$. We have not plotted higher order modes because their contribution to the torque can be neglected. In Fig. \ref{fig:sp}, we can thus compare the obtained torques with single \textit{absorption} resonances, since the resonances of $A_j$, $\vert a_j\vert $ and $\mathrm{Re} \left\lbrace a_j  \right\rbrace$ happen at different spectral positions \cite{Makitalo2014}.

We observe in Fig. \ref{fig:sp} that most of the peaks in $N_y(\lambda)$ are due to single absorption resonances of the particle. We see that the most prominent resonances for both $\phat$ and $\shat$ waves are the electric (EQ) and magnetic quadrupolar (MQ) resonance, respectively. Furthermore, absorption magnetic dipolar (MD) and magnetic octupolar (MO) resonances can also be singled out, both of them being excited by the $\shat$ evanescent wave. These observations are consistent with the two main features of the expansion of TE and TM evanescent plane waves. The first has been shown in Fig. \ref{fig:jyvsjz}: The projection of evanescent waves with multipolar modes grows exponentially with $j$ and $m_y$, reaching a maximum when $m_y=\pm j$. This enhances the role of higher multipolar orders such as the quadrupole. The second one is illustrated in Tab. \ref{tab:tetm}: When $m_y=\pm j$, a $\phat$ ($\shat$) wave only excites modes of the electric (magnetic) parity. Furthermore, looking at Eq. (\ref{eq:Nzmy}), we see that the modes with the largest $m_y$ are those that contribute most to the torque. Putting all this together, we can now figure out why TM(TE) evanescent plane waves are able to efficiently excite higher order multipolar resonances of the electric (magnetic) parity with their maximum possible transverse angular momentum value. Efficiency here must be understood as relative to excitation of lower order resonances. We note that the situation is quite different when engineering the multipolar content of a propagating beam \cite{Zambrana2012,ZambranaThesis}, where the exponential growth does not happen. 
 Notably, the polarization dependent excitation of multipolar absorption resonances does not qualitatively depend on the approximation made in Sec. \ref{sec:mietheory}. The reason is that, under a TE(TM) illumination, the secondary interaction-reflection events between the sphere and the interface will still sum up to a field that is an eigenstate of the mirror reflection across the XZ plane. This happens because the entire system, excitation-interface-sphere, has this symmetry. As argued at the end of App. \ref{sec:AppC}, the polarization controlled selection rule illustrated in Tab. \ref{tab:tetm} applies not only to plane waves, but also to general eigenstates of this particular mirror reflection.

Two last comments are in order. First, notice that, given a wavelength, the product $A_j \vert \beta_{j,m_y}^{(e)} \vert^2 $ tends to 0 for growing $j$. That is, the decay of $A_j(B_j)$ with $j$ is faster than the exponential growth of the evanescent wave coefficients $\beta_{j,m_y}^{(e)}(\beta_{j,m_y}^{(m)})$ with $j$. These conditions keep the additions in Eqs. (\ref{eq:Nxmz}-\ref{eq:Nzmy}) finite. And last but not least, as we observe in Figs. \ref{fig:angle}-\ref{fig:sp}, even though the dipolar absorption resonances of the sphere are larger than the quadrupolar ones, the torque due to evanescent plane waves turns out to be dominated by the quadrupolar orders. Unlike in many other light-matter interaction situations, where the dipolar contribution of the sample predominates, the evanescent light-matter interaction changes this pattern.

\section{Conclusion}
The decomposition of an evanescent plane wave into multipoles of well-defined transverse angular momentum has been shown to be a valuable approach to study and engineer evanescent light-matter interactions.

We have shown that when this approach is applied to experiments involving evanescent light-matter couplings, it leads to their qualitative understanding and to the ability of making rigorous quantitative predictions about them. We have also shown how the approach can be used to study the transverse torque exerted by an evanescent plane wave onto a given spherical absorbing particle, and to find
the parameters that maximize it; that is, the optimal combination of frequency, angle of incidence, and polarization of the corresponding propagating plane wave undergoing total internal reflection.

The physical insights that we have obtained can be traced back to the two main features of the decomposition of evanescent plane waves into transverse multipolar modes: A polarization independent exponential dominance of modes with large transverse angular momentum, and a polarization controlled parity selection rule.

We have provided the derivations needed to implement the transverse multipolar decomposition of an arbitrary evanescent plane wave, thereby allowing the corresponding decomposition of any evanescent field, such as, for example the field around a tapered fiber. 

We believe that the presented approach can be of help in the study and engineering of evanescent light-matter interactions.
\begin{acknowledgments}
I.F-C wishes to warmly thank Ms. Magda Felo for her help with the figures, in particular for drawing Fig. 1. We acknowledge partial financial support by the Deutsche Forschungsgemeinschaft through CRC 1173, and by the A*MIDEX project (No. ANR-11-IDEX-0001-02) funded by the Investissements d'Avenir French Government program and managed by the French National Research Agency (ANR). 
\end{acknowledgments}

\clearpage

\appendix
\section{Decomposition of an evanescent plane wave into modes with well-defined transverse angular momentum\label{sec:jymodes}}

Our goal is to obtain the decomposition of the evanescent plane wave in the top medium of Fig. \ref{fig:system_nosphere} into modes with well-defined $J_y$, and with the origin of coordinates at a distance $d$ above the interface. The expansion shall read 
\begin{equation}
	\label{eq:goal}
	\evfield=\sum_{\substack{j>0,\\|m_y|\le j}} \alpha_{j,m_y}^+\helplusmy+\alpha_{j,m_y}^-\helminusmy.
\end{equation}
The procedure that we will follow is depicted in Fig. \ref{fig:transformations}. First, we will obtain the expansion of the plane wave in a basis of multipolar fields with well-defined $J_z$ and origin in the interface [Fig. \ref{fig:transformations}(a)]. We will then shift the origin by a distance $d$ above the interface [Fig. \ref{fig:transformations}(b)], and finally we will rotate the coordinate system to obtain modes of well-defined $J_y$ at the shifted origin [Fig. \ref{fig:transformations}(c)]. Other procedures are possible. We have chosen this route because of the availability of explicit formulas for translating and rotating multipolar fields when the quantization axis for angular momentum is the $\zhat$ axis \cite{Tung1985}. Additionally, the choice of helicity ($\lambda=\pm1$) for characterizing the polarization simplifies the translation formulas, as we explain later.

Before starting, we consider the following result \cite[Eq. 7.3-15]{Tung1985}:
\begin{equation}
	\label{eq:Djm}
	\begin{split}
		\langle \mzbar\ j| R\left(\alpha,\beta,\gamma \right) |j \ \mz\rangle&= D^j\left(\alpha,\beta,\gamma\right)^{\mzbar}_{\mz}=\\\exp(-i\mzbar\alpha)d^j\left(\beta\right)^{\mzbar}_{\mz} & \exp(-i\mz\gamma),
	\end{split}
\end{equation}
where $d^j\left(\beta\right)^{\mzbar}_{\mz}$ are the reduced Wigner matrices. Equation (\ref{eq:Djm}) is the matrix element of the rotation with Euler angles $(\alpha,\beta,\gamma)$ between the two states $|j \ \mz\rangle$ and $|j \ \mbar\rangle$. The publicly available EasySpin code \cite{Stoll2006} includes a convenient implementation\footnote{See the documentation of the EasySpin function {\em wignerd}($j$,$[\alpha,\beta,\gamma]$).} of Eq. (\ref{eq:Djm}).

We now write an equation that we will use several times:
\begin{equation}
\label{eq:bible}
	\begin{split}
		&\langle \bar{\lambda}\ \mzbar\ \bar{j}\ k| R\left(\alpha,\beta,\gamma \right) \rmz=\\
		&\delta_{\bar{\lambda}\lambda}\delta_{\bar{j}j}\exp(-i\mzbar\alpha)d^j\left(\beta\right)^{\mzbar}_{\mz} \exp(-i\mz\gamma),
	\end{split}
\end{equation}
where $\delta_{nm}$ is the Kronecker delta. To write Eq. (\ref{eq:bible}), we have used that rotations do not change the eigenvalues of the linear momentum squared (equal to $k^2$), the angular momentum squared [equal to $j(j+1)$], or helicity.

Any evanescent plane wave in the top medium of Fig. (\ref{fig:system_nosphere}) can be written as the sum of two evanescent plane waves of well-defined helicity:
\begin{equation}
	\label{eq:evfieldp}
\evfield=\beta_+|\pp\ +\rangle+\beta_-|\pp\ -\rangle.
\end{equation}
For a given polarization of the impinging plane wave in the bottom medium, the coefficients $\beta_{\pm}$ in Eq. (\ref{eq:evfieldp}) can be obtained using the Fresnel formulas and the relationship between plane waves of well-defined helicity and the TE and TM plane waves \cite[Sec 2.2.3]{FerCorTHESIS}:
\begin{equation}
	\begin{split}
		\sqrt{2}|\pp \ \text{te}\rangle&=|\pp\ +\rangle+|\pp\ -\rangle\\
		\sqrt{2}|\pp \ \text{tm}\rangle&=|\pp\ +\rangle-|\pp\ -\rangle.\\
	\end{split}
\end{equation}

We start by writing a decomposition of an evanescent plane wave $|\pp \ \lambda\rangle$ into multipoles of well-defined $J_z$ and origin at the interface [Fig. \ref{fig:transformations}(a)]. We note that the typical expression of the Fresnel transmission coefficients implicitly contains the assumption that the origin of coordinates is at the interface between the two media. Objects corresponding to this origin of coordinates will be indicated by $o$ prescripts and subscripts:
\begin{equation}
	\label{eq:pl}
	|\pp \ \lambda\rangle =\sum_{\substack{j>0,\\|m_z|\le j}} \prescript{o}{}{\alpha}_{j,m_z}^\lambda\rmzo,
\end{equation}
where $k=\omega\sqrt{\epsilon_t\mu_t}$. The complex amplitudes $\prescript{o}{}{\alpha}_{j,m_z}^\lambda$ are given by the projection of each of the modes onto the plane wave, i.e. the scalar product:
\begin{equation}
	\label{eq:pwmz}
	\prescript{o}{}{\alpha}_{j,m_z}^\lambda=\prescript{}{o}{\langle} \lambda \ m_z\ j\ k|\pp \ \lambda\rangle.
\end{equation}

To obtain the complex amplitudes, we first use the following result \cite[Eq. 9.4-8]{Tung1985}: The plane wave with wavenumber $k=\omega\sqrt{\epsilon\mu}$, helicity $\lambda$, and momentum
\begin{equation}
	\pp=k\left[\sin(\theta)\cos(\phi),\sin(\theta)\sin(\phi),\cos(\theta)\right],
\end{equation}
can be obtained by rotating a plane wave of the same wavenumber and helicity with momentum aligned with the $\zhat$ axis. 
\begin{equation}
	\label{eq:prot}
	|\pp \ \lambda\rangle = R(\phi,\theta,0)|k\zhat\ \lambda\rangle.
\end{equation}
This result is valid for both propagating and evanescent plane waves.

We now use the decomposition of $|k\zhat\ \lambda\rangle$ into multipoles of well-defined $J_z$ \cite[Eq. 8.1-11]{Tung1985}:
\begin{equation}
	\label{eq:k}
	|k\zhat\ \lambda\rangle = \sum_{\oj>1}|k \ \oj\ (m_z=\lambda) \ \lambda\rangle_o,
\end{equation}
and obtain, from Eqs. (\ref{eq:pwmz}), (\ref{eq:prot}), and (\ref{eq:k})
\begin{equation}
	\label{eq:alphapwmz}
	\begin{split}
		\prescript{o}{}{\alpha}_{j,m_z}^\lambda &= \sum_{\oj>1} \langle \lambda \ m_z\ j\ k| R(\phi,\theta,0)|k\ \oj\ (m_z=\lambda) \ \lambda\rangle\\
			&\duetoref{eq:bible}\exp(-im_z\phi)d^j(\theta)^{m_z}_\lambda.
	\end{split}
\end{equation}
The last equality follows form Eq. (\ref{eq:bible}), which we indicate by $\duetoref{eq:bible}$. We will use this notation often.

Equations (\ref{eq:pl}) and (\ref{eq:alphapwmz}) provide the decomposition of the plane wave into modes of well-defined $J_z$ with the origin of coordinates at the interface between the two media.  

The next step in our program is to obtain the expansion of the same plane wave into multipoles of well-defined $J_z$ but with the origin of coordinates at a distance $d$ above the interface [Fig. \ref{fig:transformations}(b)]:

\begin{equation}
	\label{eq:aiai}
	|\pp \ \lambda\rangle =\sum_{\substack{\oj>0,\\|\mzbar|\le \oj}} {\alpha}_{\oj,\mzbar}^\lambda|k \ \oj\ \mzbar\ \lambda\rangle.
\end{equation}

The absence of the ``$o$'' prescripts and subscripts denotes objects referred to the shifted origin.
 
In order to obtain the coefficients ${\alpha}_{\oj,\mzbar}^\lambda$ as a function of the already known $\prescript{o}{}{\alpha}_{j,m_z}^\lambda$ we proceed as follows. First, since the origin shift is essentially a change of basis which does not modify the physical state, we can write:
\begin{equation}
	\label{eq:eqshift}
	\sum_{\substack{\oj>0,\\|\mzbar|\le \oj}} {\alpha}_{\oj,\mzbar}^\lambda|k \ \oj\ \mzbar\ \lambda\rangle=\sum_{\substack{j>0,\\|m_z|\le j}} \prescript{o}{}{\alpha}_{j,m_z}^\lambda\rmzo.
\end{equation}
Now, because the multipolar fields with respect to each origin form a complete basis, we may expand $\rmzo$ as a function of the $\rmz$. A crucial point here is that translations do not change helicity: 

\begin{equation}
	\label{eq:trans}
	\rmzo=\sum_{\tilde{j},\tilde{m}_z}\gamma_{j,m_z}^{{\tilde{j},\tilde{m}_z}}|k\ \tilde{j}\ \tilde{m}_z \ \lambda\rangle.
\end{equation}
Then, using the orthogonality of the multipolar fields:
\begin{equation}
	\label{eq:almost}
	{\alpha}_{\oj,\mzbar}^\lambda=\lmzbar \pp\ \lambda\rangle\duetoreftwo{eq:eqshift}{eq:trans}\sum_{\substack{j>0,\\|m_z|\le j}} \prescript{o}{}{\alpha}_{j,m_z}^\lambda\gamma_{j,m_z}^{\oj,\mzbar}.
\end{equation}
All that is left is to specify the coefficients $\gamma_{j,m_z}^{{\oj,\mzbar}}$. As seen in Eq. (\ref{eq:almost}), the $\gamma_{j,m_z}^{\oj,\mzbar}$ change the coordinates from a basis of multipoles centered at the interface to a basis of multipoles centered at $z=d$. Since we have translated the basis vectors by $d\zhat$ and the coordinates change with the inverse transformation, the $\gamma_{j,m_z}^{\oj,\mzbar}$ must be the matrix elements of a translation by $-d\zhat$ in the multipolar basis. Their analytical expression can be obtained from \cite[Prob. 9.4]{Tung1985} using that translations along the $z$ axis do not change the eigenvalue of $J_z$:
\begin{equation}
	\label{eq:do}
	\begin{split}
		&\gamma_{j,m_z}^{\oj,\mzbar}(k,\lambda)=\lmzbar T_z(-d) \rmz=\\	
		&\delta_{\mzbar\mz}\sum_{l\ge 0} (2l+1)i^lj_l(kd)\left[\mz 0(\oj l)j\mz\right]\left[j \lambda (\oj l)\lambda 0\right],
	\end{split}
\end{equation}
where $j_l(\cdot)$ is the spherical Bessel function of order $l$, $T_z(\cdot)$ is a translation along the $z$ axis, and the symbols in square brackets are Clebsch-Gordan coefficients \cite[p. 120]{Tung1985}. The publicly available EasySpin code \cite{Stoll2006} includes a convenient implementation of the Clebsch-Gordan coefficients\footnote{The $[mm'(jj')JM]$ coefficient in our notation is obtained by calling the EasySpin function {\em clebshgordan}($j,j',J,m,m',M$).} 

Equation (\ref{eq:do}) is the translation theorem for multipolar fields of well-defined helicity. Since helicity commutes with translations, the translated version of an eigenstate of helicity is still an eigenstate of helicity with the same eigenvalue. This is a simplification with respect to the case of eigenstates of parity which mix upon translation. 

Here is an important note for the use of the translation theorem in numerical calculations. The translation of a $(j,m)$ mode produces modes with $\oj=j$, but also modes with $\oj\neq j$ whose amplitudes decay with $|j-\oj|$. Conversely, a translated $\oj$ mode contains contributions from original $j\neq \oj$ modes. It is important that enough contributions of $j\neq \oj$ modes are considered from both $\oj>j$ and $\oj<j$ ``sides''. This means that for a desired maximum $\oj=\bar{j}_{max}$ value after the shift, the original expansion needs to contain enough modes with $j>\bar{j}_{max}$.

The last step is to obtain the expansion into modes of well-defined $J_y$ at the shifted origin: 
\begin{equation}
	\label{eq:joe}
	|\pp \ \lambda\rangle =\sum_{\substack{j>0,\\|m_y|\le j}} {\alpha}_{j,m_y}^\lambda|k \ j\ m_y \ \lambda\rangle.
\end{equation}

This can be achieved by a rotation of the shifted coordinate system by $-\pi/2$ around the $\xhat$ axis, as indicated in Fig. \ref{fig:transformations}(c). Since the components of angular momentum transform as a vector upon rotations, the $ R_{\xhat}(-\pi/2)$ rotation of an eigenstate of $J_z$ ($|k\ j\ m_z\ \lambda\rangle$) results in an eigenstate of $J_y$ with eigenvalue $m_y$ numerically equal to $m_z$:
\begin{equation}
	\label{eq:mymz}
	|k\ j\ (m_y=m_z)\ \lambda\rangle = R_{\xhat}(-\pi/2)|k\ j\ m_z\ \lambda\rangle.
\end{equation}
We hence have

\begin{equation}
	\label{eq:end}
	\begin{split}
		&\alpha_{j,m_y}^\lambda=\langle \ \lambda \ m_y\ j\ k|\pp\ \lambda\rangle\\
		&\duetoref{eq:mymz}\langle \lambda \ (\mzbar=m_y) \ j\ k| R_{\xhat}(\pi/2)|\pp\ \lambda\rangle\\
		&\duetoref{eq:aiai}\sum_{\substack{j>0,\\|m_z|\le j}} \alpha_{j,m_z}^\lambda \langle \ \lambda \ (\mzbar=m_y)\ j\ k| R_{\xhat}(\pi/2)|k \ j\ m_z\ \lambda\rangle\\
		&=\sum_{\substack{j>0,\\|m_z|\le j}} \alpha_{j,m_z}^\lambda \langle \ \lambda \ (\mzbar=m_y)\ j\ k| R(-\pi/2,\pi/2,\pi/2)|k \ j\ m_z\ \lambda\rangle\\
	 &\duetoref{eq:bible}\sum_{\substack{j>0,\\|m_z|\le j}} \alpha_{j,m_z}^\lambda \exp\left(im_y\pi/2\right)d^j(\pi/2)_{m_z}^{m_y}\exp(-im_z\pi/2),
	\end{split}
\end{equation}
where the fourth equality follows from the Euler angle expression of the rotation 
\begin{equation}
R_{\xhat}(\pi/2)=R(\alpha=-\pi/2,\beta=\pi/2,\gamma=\pi/2).
\end{equation}

The coefficients in Eq. (\ref{eq:goal}) can finally be obtained from Eqs. (\ref{eq:evfieldp}), (\ref{eq:joe}) and (\ref{eq:end}).

\section{Distribution of the evanescent plane wave power across multipolar subspaces\label{sec:mj}}
The decomposition of a plane wave into modes of well-defined transverse angular momentum in App. \ref{sec:jymodes} is valid for both propagating and evanescent plane waves. The only qualitative difference is the angle $\theta$ in the reduced Wigner matrix elements $d^j(\theta)_\lambda^{m_z}$ in Eq. (\ref{eq:alphapwmz}
): $\theta$ is real for propagating plane waves and complex for evanescent plane waves. This difference has a substantial effect on the distribution of the power of the plane wave across multipolar modes with different $j$.

Figure \ref{fig:djplot} is a plot of the following quantity referred to Eq. (\ref{eq:alphapwmz}):
\begin{equation}
	\label{eq:qj}
	\begin{split}
	Q_j(\theta_{in})&=\sum_{m=-j}^{m=j}\left|\prescript{o}{}{\alpha}_{j,m}^{\lambda=1} \right|^2\\
	&=\sum_{m=-j}^{m=j}\left|d^j\left[\arcsin\left(\sqrt{\frac{\epsilon_b\mu_b}{\epsilon_t\mu_t}}\sin\theta_{in}\right)\right]_{\ \ \ \lambda=1}^m\right|^2.
	\end{split}
\end{equation}
$Q_j(\theta_{in})$ measures the power of the plane wave in the top medium of Fig. (\ref{fig:system_nosphere}) contained in the $j$-th multipolar subspace as a function of the angle of incidence of the plane wave in the bottom medium. 

The behavior mentioned in the main text is evident in Fig. \ref{fig:djplot}. While the plane wave in top medium is propagating, each $j$ subspace contains exactly the same power (note the logarithmic scale in the graph). The plane wave in the top medium becomes evanescent in the shaded region. As soon as $\theta_{in}$ is larger than the angle of total internal reflection, the power is unevenly split between the different $j$ subspaces, and it concentrates exponentially at larger $j$ values. The graph is identical for $\lambda=-1$ in Eq. (\ref{eq:qj}).

\section{Restrictions on the transverse angular momentum multipolar coefficients of TE and TM plane waves\label{sec:AppC}}
The TE/TM plane waves with momentum $\pp=[p_x,0,p_z]$, defined as
\begin{equation}
	\label{eq:eig}
	\begin{split}
		\sqrt{2}|\pp \ \text{te}\rangle&=|\pp\ +\rangle+|\pp\ -\rangle,\\
		\sqrt{2}|\pp \ \text{tm}\rangle&=|\pp\ +\rangle-|\pp\ -\rangle,\\
	\end{split}
\end{equation}
are eigenstates of the mirror reflection across the XZ plane, which we denote as $\My$:
\begin{equation}
	\label{eq:eigv}
	\My|\pp \ \text{te}\rangle=-|\pp \ \text{te}\rangle,\ \My|\pp \ \text{tm}\rangle=|\pp \ \text{tm}\rangle.
\end{equation}

This imposes restrictions on the complex amplitudes of the expansion of the plane waves in multipolar fields of well-defined transverse momentum ($J_y$). To obtain the explicit form of the restrictions, we start by applying the mirror reflection $\My$ to the expansion of $|\pp\ \lambda\rangle$ in Eq. (\ref{eq:joe}):

\begin{equation}
	\label{eq:myl}
	\begin{split}
		&\My|\pp\ \lambda\rangle= \My \sum_{\substack{j>0,\\|m_y|\le j}}\alpha_{j,m_y}^\lambda \helmy\\
		&=\sum_{\substack{j>0,\\|m_y|\le j}}\alpha_{j,m_y}^\lambda \Pi R_{\yhat}(\pi)\helmy\\
		&=\sum_{\substack{j>0,\\|m_y|\le j}}\alpha_{j,m_y}^\lambda (-1)^{m_y}\Pi\helmy\\
		&=\sum_{\substack{j>0,\\|m_y|\le j}}\alpha_{j,m_y}^\lambda (-1)^{j+m_y}|k\ j\ m_y\ -\lambda\rangle,
	\end{split}
\end{equation}
where the second equality follows from $\My=\Pi R_{\yhat}(\pi)$, where $\Pi$ is the parity operator, the third equality follows from $R_{\yhat}(\pi)\helmy=\exp(-im_y\pi)\helmy$, and the fourth from \cite[Eq. 11.4-5]{Tung1985}.

Let us now focus on the expansion of the TE plane wave
\begin{equation}
\label{eq:c4}
\sqrt{2}|\pp \ \text{te}\rangle=\sum_{\substack{j>0,\\|m_y|\le j}}\alpha_{j,m_y}^+ \helplusmy+\alpha_{j,m_y}^- \helminusmy,
\end{equation}
where we apply the mirror reflection:
\begin{equation}
	\label{eq:aiaiai}
	\begin{split}
	&\sqrt{2}\My|\pp \ \text{te}\rangle\\
	&=\sum_{\substack{j>0,\\|m_y|\le j}}\alpha_{j,m_y}^+ \My\helplusmy+\alpha_{j,m_y}^-\My \helminusmy\\
	&\duetoref{eq:myl}\sum_{\substack{j>0,\\|m_y|\le j}}\alpha_{j,m_y}^+(-1)^{j+m_y}\helminusmy+\\
	&\hspace{2.375cm}\alpha_{j,m_y}^-(-1)^{j+m_y}\helplusmy.
	\end{split}
\end{equation}

Since $\My|\pp \ \text{te}\rangle=-|\pp \ \text{te}\rangle$ [Eq. (\ref{eq:eigv})], it must be that 

\begin{equation}
	\begin{split}
		&\sqrt{2}|\pp \ \text{te}\rangle\duetoref{eq:c4}\sum_{\substack{j>0,\\|m_y|\le j}}\alpha_{j,m_y}^+ \helplusmy+\alpha_{j,m_y}^- \helminusmy\\
	&\duetoref{eq:eigv}-\sqrt{2}\My|\pp \ \text{te}\rangle\\
	&\duetoref{eq:aiaiai}\sum_{\substack{j>0,\\|m_y|\le j}}-\alpha_{j,m_y}^+(-1)^{j+m_y}\helminusmy+\\
	&\hspace{2.375cm}-\alpha_{j,m_y}^-(-1)^{j+m_y}\helplusmy,
	\end{split}
\end{equation}

which forces: 
\begin{equation}
	\label{eq:rest}
	\boxed{\alpha_{j,m_y}^{\pm}=-(-1)^{j+m_y}\alpha_{j,m_y}^{\mp}}.
\end{equation}
The same result is obtained starting with the TM plane wave.

Let us now consider the multipolar basis of fields of well-defined parity \cite[Eq. 11.4-6]{Tung1985}:
\begin{equation}
	\label{eq:hp}
	\begin{split}
		\sqrt{2}|k\ j\ m_y\ \tau=1\rangle &= |k\ j\ m_y\ +\rangle+|k\ j\ m_y\ -\rangle,\\
		\sqrt{2}|k\ j\ m_y\ \tau=-1\rangle &= |k\ j\ m_y\ +\rangle-|k\ j\ m_y\ -\rangle,
	\end{split}
\end{equation}
where the parity of $|k\ j\ m_y\ \tau\rangle$ is $\tau(-1)^j$ , and $\tau=1(-1)$ corresponds to the ``electric''(``magnetic'') multipoles \cite[Eq. 11.4-25]{Tung1985}.

Using Eq. (\ref{eq:hp}) we can establish the relationship between the multipolar helicity coefficients $\alpha_{j,m_y}^+$ and the multipolar parity coefficients $\prescript{+}{}{\beta}_{j,m_y}^{\tau}$ of the two multipolar expansions of $|\pp\ +\rangle$:
\begin{equation}
	\label{eq:betas}
	\prescript{+}{}{\beta}_{j,m_y}^{\tau=1}=\prescript{+}{}{\beta}_{j,m_y}^{\tau=-1}=\frac{\alpha_{j,m_y}^+}{\sqrt{2}}.
\end{equation}
Parallel steps for $|\pp\ -\rangle$ result in
\begin{equation}
	\label{eq:betap}
	\prescript{-}{}{\beta}_{j,m_y}^{\tau=1}=-\left(\prescript{-}{}{\beta}_{j,m_y}^{\tau=-1}\right)=\frac{\alpha_{j,m_y}^-}{\sqrt{2}}.
\end{equation}

Equations (\ref{eq:eig}), (\ref{eq:c4}), (\ref{eq:betas}), and (\ref{eq:betap}) allow us to now write:
\begin{widetext}
\begin{equation}
	\label{eq:taus}
	\begin{split}
		\sqrt{2}|\pp \ te\rangle&=\sum_{\substack{j>0,\\|m_y|\le j}} \left(\alpha_{j,m_y}^++\alpha_{j,m_y}^-\right)|k\ j\ m_y\ \tau=1\rangle+ \left(\alpha_{j,m_y}^+-\alpha_{j,m_y}^-\right)|k\ j\ m_y\ \tau=-1\rangle,\\
		\sqrt{2}|\pp \ tm\rangle&=\sum_{\substack{j>0,\\|m_y|\le j}} \left(\alpha_{j,m_y}^+-\alpha_{j,m_y}^-\right)|k\ j\ m_y\ \tau=1\rangle+ \left(\alpha_{j,m_y}^++\alpha_{j,m_y}^-\right)|k\ j\ m_y\ \tau=-1\rangle.\\
	\end{split}
\end{equation}
\end{widetext}

Finally, restriction (\ref{eq:rest}) applied to Eq. (\ref{eq:taus}) results in

\begin{widetext}
	\begin{ChangeMargin}{-10.0em}{-20.0ex}
	\begin{empheq}[box=\fbox]{align}
		\label{eq:result}
	\nonumber
	\sqrt{2}|\pp \ te\rangle&=\sum_{\substack{j>0,\\|m_y|\le j}} \alpha_{j,m_y}^+\left\{\left[\shadedbox{1-(-1)^{j+m_y}}\right]|k\ j\ m_y\ \tau=1\rangle+ \left[\shadedbox{1+(-1)^{j+m_y}}\right]|k\ j\ m_y\ \tau=-1\rangle\right\},\\
	\sqrt{2}|\pp \ tm\rangle&=\sum_{\substack{j>0,\\|m_y|\le j}} \alpha_{j,m_y}^+\left\{\left[\shadedbox{1+(-1)^{j+m_y}}\right]|k\ j\ m_y\ \tau=1\rangle+ \left[\shadedbox{1-(-1)^{j+m_y}}\right]|k\ j\ m_y\ \tau=-1\rangle\right\}.
\end{empheq}
\end{ChangeMargin}
\end{widetext}

Equation (\ref{eq:result}) contains the following parity selection rule: Given a $(j,m_y)$ pair, the TE(TM) polarized plane waves have a contribution from only one of the two possible parities, and the contributing parity is opposite for the TE versus the TM polarized plane wave.

It should be noted that, in a transmission or reflection problem, the TE(TM) plane waves in Eq. (\ref{eq:result}) would be multiplied by the corresponding Fresnel coefficients.

An important generalization of the result in Eq. (\ref{eq:result}) is possible. It is straightforward to check that one can substitute the TE and TM plane waves by general eigenstates of $M_{\yhat}$ with eigenvalues -1 and +1, respectively, and all the steps in the derivations remain unchanged. In essence, we have only used the relationships between multipolar fields of well-defined helicity and multipolar fields of well-defined parity, and the fact that the fields are eigenstates of $M_{\yhat}$.


\begin{thebibliography}{53}\makeatletter
\providecommand \@ifxundefined [1]{ \@ifx{#1\undefined}
}\providecommand \@ifnum [1]{ \ifnum #1\expandafter \@firstoftwo
 \else \expandafter \@secondoftwo
 \fi
}\providecommand \@ifx [1]{ \ifx #1\expandafter \@firstoftwo
 \else \expandafter \@secondoftwo
 \fi
}\providecommand \natexlab [1]{#1}\providecommand \enquote  [1]{``#1''}\providecommand \bibnamefont  [1]{#1}\providecommand \bibfnamefont [1]{#1}\providecommand \citenamefont [1]{#1}\providecommand \href@noop [0]{\@secondoftwo}\providecommand \href [0]{\begingroup \@sanitize@url \@href}\providecommand \@href[1]{\@@startlink{#1}\@@href}\providecommand \@@href[1]{\endgroup#1\@@endlink}\providecommand \@sanitize@url [0]{\catcode `\\12\catcode `\$12\catcode
  `\&12\catcode `\#12\catcode `\^12\catcode `\_12\catcode `\%12\relax}\providecommand \@@startlink[1]{}\providecommand \@@endlink[0]{}\providecommand \url  [0]{\begingroup\@sanitize@url \@url }\providecommand \@url [1]{\endgroup\@href {#1}{\urlprefix }}\providecommand \urlprefix  [0]{URL }\providecommand \Eprint [0]{\href }\providecommand \doibase [0]{http://dx.doi.org/}\providecommand \selectlanguage [0]{\@gobble}\providecommand \bibinfo  [0]{\@secondoftwo}\providecommand \bibfield  [0]{\@secondoftwo}\providecommand \translation [1]{[#1]}\providecommand \BibitemOpen [0]{}\providecommand \bibitemStop [0]{}\providecommand \bibitemNoStop [0]{.\EOS\space}\providecommand \EOS [0]{\spacefactor3000\relax}\providecommand \BibitemShut  [1]{\csname bibitem#1\endcsname}\let\auto@bib@innerbib\@empty
\bibitem [{\citenamefont {Mabuchi}\ and\ \citenamefont
  {Kimble}(1994)}]{Mabuchi1994}  \BibitemOpen
  \bibfield  {author} {\bibinfo {author} {\bibfnamefont {H.}~\bibnamefont
  {Mabuchi}}\ and\ \bibinfo {author} {\bibfnamefont {H.~J.}\ \bibnamefont
  {Kimble}},\ }\href {\doibase 10.1364/OL.19.000749} {\bibfield  {journal}
  {\bibinfo  {journal} {Opt. Lett.}\ }\textbf {\bibinfo {volume} {19}},\
  \bibinfo {pages} {749} (\bibinfo {year} {1994})}\BibitemShut {NoStop}\bibitem [{\citenamefont {van Enk}\ and\ \citenamefont
  {Nienhuis}(1994)}]{Van1994}  \BibitemOpen
  \bibfield  {author} {\bibinfo {author} {\bibfnamefont {S.}~\bibnamefont {van
  Enk}}\ and\ \bibinfo {author} {\bibfnamefont {G.}~\bibnamefont {Nienhuis}},\
  }\href {\doibase http://dx.doi.org/10.1016/0030-4018(94)00461-7} {\bibfield
  {journal} {\bibinfo  {journal} {Opt. Comm.}\ }\textbf {\bibinfo {volume}
  {112}},\ \bibinfo {pages} {225 } (\bibinfo {year} {1994})}\BibitemShut
  {NoStop}\bibitem [{\citenamefont {Novotny}(1996)}]{Novotny1996}  \BibitemOpen
  \bibfield  {author} {\bibinfo {author} {\bibfnamefont {L.}~\bibnamefont
  {Novotny}},\ }\href {\doibase 10.1063/1.117111} {\bibfield  {journal}
  {\bibinfo  {journal} {Applied Physics Letters}\ }\textbf {\bibinfo {volume}
  {69}},\ \bibinfo {pages} {3806–3808} (\bibinfo {year} {1996})}\BibitemShut
  {NoStop}\bibitem [{\citenamefont {Tojo}\ \emph {et~al.}(2004)\citenamefont {Tojo},
  \citenamefont {Hasuo},\ and\ \citenamefont {Fujimoto}}]{Tojo2004}  \BibitemOpen
  \bibfield  {author} {\bibinfo {author} {\bibfnamefont {S.}~\bibnamefont
  {Tojo}}, \bibinfo {author} {\bibfnamefont {M.}~\bibnamefont {Hasuo}}, \ and\
  \bibinfo {author} {\bibfnamefont {T.}~\bibnamefont {Fujimoto}},\ }\href
  {\doibase 10.1103/PhysRevLett.92.053001} {\bibfield  {journal} {\bibinfo
  {journal} {Phys. Rev. Lett.}\ }\textbf {\bibinfo {volume} {92}},\ \bibinfo
  {pages} {053001} (\bibinfo {year} {2004})}\BibitemShut {NoStop}\bibitem [{\citenamefont {Tojo}\ and\ \citenamefont {Hasuo}(2005)}]{Tojo2005}  \BibitemOpen
  \bibfield  {author} {\bibinfo {author} {\bibfnamefont {S.}~\bibnamefont
  {Tojo}}\ and\ \bibinfo {author} {\bibfnamefont {M.}~\bibnamefont {Hasuo}},\
  }\href@noop {} {\bibfield  {journal} {\bibinfo  {journal} {Phys. Rev. A}\
  }\textbf {\bibinfo {volume} {71}},\ \bibinfo {pages} {012508} (\bibinfo
  {year} {2005})}\BibitemShut {NoStop}\bibitem [{\citenamefont {Bharadwaj}\ and\ \citenamefont
  {Novotny}(2007)}]{Bharadwaj2007}  \BibitemOpen
  \bibfield  {author} {\bibinfo {author} {\bibfnamefont {P.}~\bibnamefont
  {Bharadwaj}}\ and\ \bibinfo {author} {\bibfnamefont {L.}~\bibnamefont
  {Novotny}},\ }\href {\doibase 10.1364/OE.15.014266} {\bibfield  {journal}
  {\bibinfo  {journal} {Opt. Express}\ }\textbf {\bibinfo {volume} {15}},\
  \bibinfo {pages} {14266} (\bibinfo {year} {2007})}\BibitemShut {NoStop}\bibitem [{\citenamefont {Lembessis}\ \emph {et~al.}(2009)\citenamefont
  {Lembessis}, \citenamefont {Babiker},\ and\ \citenamefont
  {Andrews}}]{Lembessis2009}  \BibitemOpen
  \bibfield  {author} {\bibinfo {author} {\bibfnamefont {V.}~\bibnamefont
  {Lembessis}}, \bibinfo {author} {\bibfnamefont {M.}~\bibnamefont {Babiker}},
  \ and\ \bibinfo {author} {\bibfnamefont {D.}~\bibnamefont {Andrews}},\
  }\href@noop {} {\bibfield  {journal} {\bibinfo  {journal} {Phys. Rev. A}\
  }\textbf {\bibinfo {volume} {79}},\ \bibinfo {pages} {011806} (\bibinfo
  {year} {2009})}\BibitemShut {NoStop}\bibitem [{\citenamefont {Andrews}(2010)}]{Andrews2010}  \BibitemOpen
  \bibfield  {author} {\bibinfo {author} {\bibfnamefont {D.~L.}\ \bibnamefont
  {Andrews}},\ }\href {\doibase 10.1103/PhysRevA.81.033825} {\bibfield
  {journal} {\bibinfo  {journal} {Phys. Rev. A}\ }\textbf {\bibinfo {volume}
  {81}},\ \bibinfo {pages} {033825} (\bibinfo {year} {2010})}\BibitemShut
  {NoStop}\bibitem [{\citenamefont {Le~Kien}\ and\ \citenamefont
  {Rauschenbeutel}(2014)}]{Le2014}  \BibitemOpen
  \bibfield  {author} {\bibinfo {author} {\bibfnamefont {F.}~\bibnamefont
  {Le~Kien}}\ and\ \bibinfo {author} {\bibfnamefont {A.}~\bibnamefont
  {Rauschenbeutel}},\ }\href@noop {} {\bibfield  {journal} {\bibinfo  {journal}
  {Phys. Rev. A}\ }\textbf {\bibinfo {volume} {90}},\ \bibinfo {pages} {023805}
  (\bibinfo {year} {2014})}\BibitemShut {NoStop}\bibitem [{\citenamefont {Reitz}\ \emph {et~al.}(2014)\citenamefont {Reitz},
  \citenamefont {Sayrin}, \citenamefont {Albrecht}, \citenamefont {Mazets},
  \citenamefont {Mitsch}, \citenamefont {Schneeweiss},\ and\ \citenamefont
  {Rauschenbeutel}}]{Reitz2014}  \BibitemOpen
  \bibfield  {author} {\bibinfo {author} {\bibfnamefont {D.}~\bibnamefont
  {Reitz}}, \bibinfo {author} {\bibfnamefont {C.}~\bibnamefont {Sayrin}},
  \bibinfo {author} {\bibfnamefont {B.}~\bibnamefont {Albrecht}}, \bibinfo
  {author} {\bibfnamefont {I.}~\bibnamefont {Mazets}}, \bibinfo {author}
  {\bibfnamefont {R.}~\bibnamefont {Mitsch}}, \bibinfo {author} {\bibfnamefont
  {P.}~\bibnamefont {Schneeweiss}}, \ and\ \bibinfo {author} {\bibfnamefont
  {A.}~\bibnamefont {Rauschenbeutel}},\ }\href@noop {} {\bibfield  {journal}
  {\bibinfo  {journal} {Phys. Rev. A}\ }\textbf {\bibinfo {volume} {89}},\
  \bibinfo {pages} {031804} (\bibinfo {year} {2014})}\BibitemShut {NoStop}\bibitem [{\citenamefont {Bliokh}\ \emph {et~al.}(2014)\citenamefont {Bliokh},
  \citenamefont {Bekshaev},\ and\ \citenamefont {Nori}}]{Bliokh2014b}  \BibitemOpen
  \bibfield  {author} {\bibinfo {author} {\bibfnamefont {K.~Y.}\ \bibnamefont
  {Bliokh}}, \bibinfo {author} {\bibfnamefont {A.~Y.}\ \bibnamefont
  {Bekshaev}}, \ and\ \bibinfo {author} {\bibfnamefont {F.}~\bibnamefont
  {Nori}},\ }\href@noop {} {\bibfield  {journal} {\bibinfo  {journal} {Nat.
  Commun.}\ }\textbf {\bibinfo {volume} {5}} (\bibinfo {year}
  {2014})}\BibitemShut {NoStop}\bibitem [{\citenamefont {Mitsch}\ \emph {et~al.}(2014)\citenamefont {Mitsch},
  \citenamefont {Sayrin}, \citenamefont {Albrecht}, \citenamefont
  {Schneeweiss},\ and\ \citenamefont {Rauschenbeutel}}]{Mitsch2014}  \BibitemOpen
  \bibfield  {author} {\bibinfo {author} {\bibfnamefont {R.}~\bibnamefont
  {Mitsch}}, \bibinfo {author} {\bibfnamefont {C.}~\bibnamefont {Sayrin}},
  \bibinfo {author} {\bibfnamefont {B.}~\bibnamefont {Albrecht}}, \bibinfo
  {author} {\bibfnamefont {P.}~\bibnamefont {Schneeweiss}}, \ and\ \bibinfo
  {author} {\bibfnamefont {A.}~\bibnamefont {Rauschenbeutel}},\ }\href@noop {}
  {\bibfield  {journal} {\bibinfo  {journal} {Phys. Rev. A}\ }\textbf {\bibinfo
  {volume} {89}},\ \bibinfo {pages} {063829} (\bibinfo {year}
  {2014})}\BibitemShut {NoStop}\bibitem [{\citenamefont {Petersen}\ \emph {et~al.}(2014)\citenamefont
  {Petersen}, \citenamefont {Volz},\ and\ \citenamefont
  {Rauschenbeutel}}]{Petersen2014}  \BibitemOpen
  \bibfield  {author} {\bibinfo {author} {\bibfnamefont {J.}~\bibnamefont
  {Petersen}}, \bibinfo {author} {\bibfnamefont {J.}~\bibnamefont {Volz}}, \
  and\ \bibinfo {author} {\bibfnamefont {A.}~\bibnamefont {Rauschenbeutel}},\
  }\href@noop {} {\bibfield  {journal} {\bibinfo  {journal} {Science}\ }\textbf
  {\bibinfo {volume} {346}},\ \bibinfo {pages} {67} (\bibinfo {year}
  {2014})}\BibitemShut {NoStop}\bibitem [{\citenamefont {Rodr\'{i}guez-Fortu\~{n}o}\ \emph
  {et~al.}(2014)\citenamefont {Rodr\'{i}guez-Fortu\~{n}o}, \citenamefont
  {Barber-Sanz}, \citenamefont {Puerto}, \citenamefont {Griol},\ and\
  \citenamefont {Mart{\'\i}nez}}]{Rodriguez2014}  \BibitemOpen
  \bibfield  {author} {\bibinfo {author} {\bibfnamefont {F.~J.}\ \bibnamefont
  {Rodr\'{i}guez-Fortu\~{n}o}}, \bibinfo {author} {\bibfnamefont
  {I.}~\bibnamefont {Barber-Sanz}}, \bibinfo {author} {\bibfnamefont
  {D.}~\bibnamefont {Puerto}}, \bibinfo {author} {\bibfnamefont
  {A.}~\bibnamefont {Griol}}, \ and\ \bibinfo {author} {\bibfnamefont
  {A.}~\bibnamefont {Mart{\'\i}nez}},\ }\href@noop {} {\bibfield  {journal}
  {\bibinfo  {journal} {ACS Photonics}\ }\textbf {\bibinfo {volume} {1}},\
  \bibinfo {pages} {762} (\bibinfo {year} {2014})}\BibitemShut {NoStop}\bibitem [{\citenamefont {S{\"o}llner}\ \emph {et~al.}(2015)\citenamefont
  {S{\"o}llner}, \citenamefont {Mahmoodian}, \citenamefont {Hansen},
  \citenamefont {Midolo}, \citenamefont {Javadi}, \citenamefont
  {Kir{\v{s}}ansk{\.e}}, \citenamefont {Pregnolato}, \citenamefont {El-Ella},
  \citenamefont {Lee}, \citenamefont {Song} \emph {et~al.}}]{Sollner2015}  \BibitemOpen
  \bibfield  {author} {\bibinfo {author} {\bibfnamefont {I.}~\bibnamefont
  {S{\"o}llner}}, \bibinfo {author} {\bibfnamefont {S.}~\bibnamefont
  {Mahmoodian}}, \bibinfo {author} {\bibfnamefont {S.~L.}\ \bibnamefont
  {Hansen}}, \bibinfo {author} {\bibfnamefont {L.}~\bibnamefont {Midolo}},
  \bibinfo {author} {\bibfnamefont {A.}~\bibnamefont {Javadi}}, \bibinfo
  {author} {\bibfnamefont {G.}~\bibnamefont {Kir{\v{s}}ansk{\.e}}}, \bibinfo
  {author} {\bibfnamefont {T.}~\bibnamefont {Pregnolato}}, \bibinfo {author}
  {\bibfnamefont {H.}~\bibnamefont {El-Ella}}, \bibinfo {author} {\bibfnamefont
  {E.~H.}\ \bibnamefont {Lee}}, \bibinfo {author} {\bibfnamefont {J.~D.}\
  \bibnamefont {Song}},  \emph {et~al.},\ }\href@noop {} {\bibfield  {journal}
  {\bibinfo  {journal} {Nat. Nanotech.}\ }\textbf {\bibinfo {volume} {10}},\
  \bibinfo {pages} {775} (\bibinfo {year} {2015})}\BibitemShut {NoStop}\bibitem [{\citenamefont {Bliokh}\ \emph {et~al.}(2015)\citenamefont {Bliokh},
  \citenamefont {Smirnova},\ and\ \citenamefont {Nori}}]{Bliokh2015}  \BibitemOpen
  \bibfield  {author} {\bibinfo {author} {\bibfnamefont {K.~Y.}\ \bibnamefont
  {Bliokh}}, \bibinfo {author} {\bibfnamefont {D.}~\bibnamefont {Smirnova}}, \
  and\ \bibinfo {author} {\bibfnamefont {F.}~\bibnamefont {Nori}},\ }\href@noop
  {} {\bibfield  {journal} {\bibinfo  {journal} {Science}\ }\textbf {\bibinfo
  {volume} {348}},\ \bibinfo {pages} {1448} (\bibinfo {year}
  {2015})}\BibitemShut {NoStop}\bibitem [{\citenamefont {Van~Mechelen}\ and\ \citenamefont
  {Jacob}(2016)}]{Van2016}  \BibitemOpen
  \bibfield  {author} {\bibinfo {author} {\bibfnamefont {T.}~\bibnamefont
  {Van~Mechelen}}\ and\ \bibinfo {author} {\bibfnamefont {Z.}~\bibnamefont
  {Jacob}},\ }\href@noop {} {\bibfield  {journal} {\bibinfo  {journal}
  {Optica}\ }\textbf {\bibinfo {volume} {3}},\ \bibinfo {pages} {118} (\bibinfo
  {year} {2016})}\BibitemShut {NoStop}\bibitem [{\citenamefont {Bliokh}\ and\ \citenamefont
  {Nori}(2012)}]{Bliokh2012}  \BibitemOpen
  \bibfield  {author} {\bibinfo {author} {\bibfnamefont {K.~Y.}\ \bibnamefont
  {Bliokh}}\ and\ \bibinfo {author} {\bibfnamefont {F.}~\bibnamefont {Nori}},\
  }\href@noop {} {\bibfield  {journal} {\bibinfo  {journal} {Phys. Rev. A}\
  }\textbf {\bibinfo {volume} {85}},\ \bibinfo {pages} {061801} (\bibinfo
  {year} {2012})}\BibitemShut {NoStop}\bibitem [{\citenamefont {Aiello}\ \emph {et~al.}(2009)\citenamefont {Aiello},
  \citenamefont {Lindlein}, \citenamefont {Marquardt},\ and\ \citenamefont
  {Leuchs}}]{Aiello2009}  \BibitemOpen
  \bibfield  {author} {\bibinfo {author} {\bibfnamefont {A.}~\bibnamefont
  {Aiello}}, \bibinfo {author} {\bibfnamefont {N.}~\bibnamefont {Lindlein}},
  \bibinfo {author} {\bibfnamefont {C.}~\bibnamefont {Marquardt}}, \ and\
  \bibinfo {author} {\bibfnamefont {G.}~\bibnamefont {Leuchs}},\ }\href
  {\doibase 10.1103/PhysRevLett.103.100401} {\bibfield  {journal} {\bibinfo
  {journal} {Phys. Rev. Lett.}\ }\textbf {\bibinfo {volume} {103}},\ \bibinfo
  {pages} {100401} (\bibinfo {year} {2009})}\BibitemShut {NoStop}\bibitem [{\citenamefont {Aiello}\ \emph {et~al.}(2015)\citenamefont {Aiello},
  \citenamefont {Banzer}, \citenamefont {Neugebauer},\ and\ \citenamefont
  {Leuchs}}]{Aiello2015}  \BibitemOpen
  \bibfield  {author} {\bibinfo {author} {\bibfnamefont {A.}~\bibnamefont
  {Aiello}}, \bibinfo {author} {\bibfnamefont {P.}~\bibnamefont {Banzer}},
  \bibinfo {author} {\bibfnamefont {M.}~\bibnamefont {Neugebauer}}, \ and\
  \bibinfo {author} {\bibfnamefont {G.}~\bibnamefont {Leuchs}},\ }\href
  {http://dx.doi.org/10.1038/nphoton.2015.203} {\bibfield  {journal} {\bibinfo
  {journal} {Nat Photon}\ }\textbf {\bibinfo {volume} {9}},\ \bibinfo {pages}
  {789} (\bibinfo {year} {2015})},\ \bibinfo {note} {progress
  Article}\BibitemShut {NoStop}\bibitem [{\citenamefont {Friese}\ \emph {et~al.}(1996)\citenamefont {Friese},
  \citenamefont {Enger}, \citenamefont {Rubinsztein-Dunlop},\ and\
  \citenamefont {Heckenberg}}]{Friese1996}  \BibitemOpen
  \bibfield  {author} {\bibinfo {author} {\bibfnamefont {M.~E.~J.}\
  \bibnamefont {Friese}}, \bibinfo {author} {\bibfnamefont {J.}~\bibnamefont
  {Enger}}, \bibinfo {author} {\bibfnamefont {H.}~\bibnamefont
  {Rubinsztein-Dunlop}}, \ and\ \bibinfo {author} {\bibfnamefont {N.~R.}\
  \bibnamefont {Heckenberg}},\ }\href {\doibase 10.1103/PhysRevA.54.1593}
  {\bibfield  {journal} {\bibinfo  {journal} {Phys. Rev. A}\ }\textbf {\bibinfo
  {volume} {54}},\ \bibinfo {pages} {1593} (\bibinfo {year}
  {1996})}\BibitemShut {NoStop}\bibitem [{\citenamefont {Simpson}\ \emph {et~al.}(1997)\citenamefont
  {Simpson}, \citenamefont {Dholakia}, \citenamefont {Allen},\ and\
  \citenamefont {Padgett}}]{Simpson1997}  \BibitemOpen
  \bibfield  {author} {\bibinfo {author} {\bibfnamefont {N.~B.}\ \bibnamefont
  {Simpson}}, \bibinfo {author} {\bibfnamefont {K.}~\bibnamefont {Dholakia}},
  \bibinfo {author} {\bibfnamefont {L.}~\bibnamefont {Allen}}, \ and\ \bibinfo
  {author} {\bibfnamefont {M.~J.}\ \bibnamefont {Padgett}},\ }\href {\doibase
  10.1364/OL.22.000052} {\bibfield  {journal} {\bibinfo  {journal} {Opt.
  Lett.}\ }\textbf {\bibinfo {volume} {22}},\ \bibinfo {pages} {52} (\bibinfo
  {year} {1997})}\BibitemShut {NoStop}\bibitem [{\citenamefont {Chew}\ \emph {et~al.}(1979)\citenamefont {Chew},
  \citenamefont {Wang},\ and\ \citenamefont {Kerker}}]{Chew1979}  \BibitemOpen
  \bibfield  {author} {\bibinfo {author} {\bibfnamefont {H.}~\bibnamefont
  {Chew}}, \bibinfo {author} {\bibfnamefont {D.-S.}\ \bibnamefont {Wang}}, \
  and\ \bibinfo {author} {\bibfnamefont {M.}~\bibnamefont {Kerker}},\
  }\href@noop {} {\bibfield  {journal} {\bibinfo  {journal} {App. Opt.}\
  }\textbf {\bibinfo {volume} {18}},\ \bibinfo {pages} {2679} (\bibinfo {year}
  {1979})}\BibitemShut {NoStop}\bibitem [{\citenamefont {Liu}\ \emph {et~al.}(1995)\citenamefont {Liu},
  \citenamefont {Kaiser}, \citenamefont {Lange},\ and\ \citenamefont
  {Schweiger}}]{Liu1995}  \BibitemOpen
  \bibfield  {author} {\bibinfo {author} {\bibfnamefont {C.}~\bibnamefont
  {Liu}}, \bibinfo {author} {\bibfnamefont {T.}~\bibnamefont {Kaiser}},
  \bibinfo {author} {\bibfnamefont {S.}~\bibnamefont {Lange}}, \ and\ \bibinfo
  {author} {\bibfnamefont {G.}~\bibnamefont {Schweiger}},\ }\href@noop {}
  {\bibfield  {journal} {\bibinfo  {journal} {Opt. Comm.}\ }\textbf {\bibinfo
  {volume} {117}},\ \bibinfo {pages} {521} (\bibinfo {year}
  {1995})}\BibitemShut {NoStop}\bibitem [{\citenamefont {Zvyagin}\ and\ \citenamefont
  {Goto}(1998)}]{Zvyagin1998}  \BibitemOpen
  \bibfield  {author} {\bibinfo {author} {\bibfnamefont {A.}~\bibnamefont
  {Zvyagin}}\ and\ \bibinfo {author} {\bibfnamefont {K.}~\bibnamefont {Goto}},\
  }\href@noop {} {\bibfield  {journal} {\bibinfo  {journal} {JOSA A}\ }\textbf
  {\bibinfo {volume} {15}},\ \bibinfo {pages} {3003} (\bibinfo {year}
  {1998})}\BibitemShut {NoStop}\bibitem [{\citenamefont {Bekshaev}\ \emph {et~al.}(2013)\citenamefont
  {Bekshaev}, \citenamefont {Bliokh},\ and\ \citenamefont
  {Nori}}]{Bekshaev2013}  \BibitemOpen
  \bibfield  {author} {\bibinfo {author} {\bibfnamefont {A.~Y.}\ \bibnamefont
  {Bekshaev}}, \bibinfo {author} {\bibfnamefont {K.~Y.}\ \bibnamefont
  {Bliokh}}, \ and\ \bibinfo {author} {\bibfnamefont {F.}~\bibnamefont
  {Nori}},\ }\href {\doibase 10.1364/OE.21.007082} {\bibfield  {journal}
  {\bibinfo  {journal} {Opt. Express}\ }\textbf {\bibinfo {volume} {21}},\
  \bibinfo {pages} {7082} (\bibinfo {year} {2013})}\BibitemShut {NoStop}\bibitem [{\citenamefont {Stoll}\ and\ \citenamefont
  {Schweiger}(2006)}]{Stoll2006}  \BibitemOpen
  \bibfield  {author} {\bibinfo {author} {\bibfnamefont {S.}~\bibnamefont
  {Stoll}}\ and\ \bibinfo {author} {\bibfnamefont {A.}~\bibnamefont
  {Schweiger}},\ }\href@noop {} {\bibfield  {journal} {\bibinfo  {journal} {J.
  Magn. Reson.}\ }\textbf {\bibinfo {volume} {178}},\ \bibinfo {pages} {42}
  (\bibinfo {year} {2006})}\BibitemShut {NoStop}\bibitem [{\citenamefont {Tung}(1985)}]{Tung1985}  \BibitemOpen
  \bibfield  {author} {\bibinfo {author} {\bibfnamefont {W.-K.}\ \bibnamefont
  {Tung}},\ }\href@noop {} {\emph {\bibinfo {title} {Group Theory in
  Physics}}}\ (\bibinfo  {publisher} {World Scientific},\ \bibinfo {year}
  {1985})\BibitemShut {NoStop}\bibitem [{\citenamefont {Berestetskii}\ \emph {et~al.}(1982)\citenamefont
  {Berestetskii}, \citenamefont {Pitaevskii},\ and\ \citenamefont
  {Lifshitz}}]{Berestetskii1982}  \BibitemOpen
  \bibfield  {author} {\bibinfo {author} {\bibfnamefont {V.~B.}\ \bibnamefont
  {Berestetskii}}, \bibinfo {author} {\bibfnamefont {L.~P.}\ \bibnamefont
  {Pitaevskii}}, \ and\ \bibinfo {author} {\bibfnamefont {E.~M.}\ \bibnamefont
  {Lifshitz}},\ }\href
  {http://www.amazon.com/exec/obidos/redirect?tag=citeulike07-20\&path=ASIN/0750633719}
  {\emph {\bibinfo {title} {{Quantum Electrodynamics, Second Edition: Volume
  4}}}},\ \bibinfo {edition} {2nd}\ ed.\ (\bibinfo  {publisher}
  {Butterworth-Heinemann},\ \bibinfo {address} {Oxford},\ \bibinfo {year}
  {1982})\BibitemShut {NoStop}\bibitem [{\citenamefont {Fernandez-Corbaton}(2014)}]{FerCorTHESIS}  \BibitemOpen
  \bibfield  {author} {\bibinfo {author} {\bibfnamefont {I.}~\bibnamefont
  {Fernandez-Corbaton}},\ }\emph {\bibinfo {title} {Helicity and duality
  symmetry in light matter interactions: Theory and applications}},\ \href
  {http://arxiv.org/abs/1407.4432} {Ph.D. thesis},\ \bibinfo  {school}
  {Macquarie University} (\bibinfo {year} {2014}),\ \bibinfo {note} {{arXiv}:
  1407.4432}\BibitemShut {NoStop}\bibitem [{\citenamefont {Fernandez-Corbaton}\ and\ \citenamefont
  {Molina-Terriza}(2013)}]{FerCor2013}  \BibitemOpen
  \bibfield  {author} {\bibinfo {author} {\bibfnamefont {I.}~\bibnamefont
  {Fernandez-Corbaton}}\ and\ \bibinfo {author} {\bibfnamefont
  {G.}~\bibnamefont {Molina-Terriza}},\ }\href@noop {} {\bibfield  {journal}
  {\bibinfo  {journal} {Phys. Rev. B}\ }\textbf {\bibinfo {volume} {88}},\
  \bibinfo {pages} {085111} (\bibinfo {year} {2013})}\BibitemShut {NoStop}\bibitem [{\citenamefont {Zambrana-Puyalto}\ and\ \citenamefont
  {Bonod}(2016)}]{Zambrana2016Nano}  \BibitemOpen
  \bibfield  {author} {\bibinfo {author} {\bibfnamefont {X.}~\bibnamefont
  {Zambrana-Puyalto}}\ and\ \bibinfo {author} {\bibfnamefont {N.}~\bibnamefont
  {Bonod}},\ }\href {\doibase 10.1039/C6NR00676K} {\bibfield  {journal}
  {\bibinfo  {journal} {Nanoscale}\ }\textbf {\bibinfo {volume} {8}},\ \bibinfo
  {pages} {10441} (\bibinfo {year} {2016})}\BibitemShut {NoStop}\bibitem [{\citenamefont {Almaas}\ and\ \citenamefont
  {Brevik}(1995)}]{Almaas1995}  \BibitemOpen
  \bibfield  {author} {\bibinfo {author} {\bibfnamefont {E.}~\bibnamefont
  {Almaas}}\ and\ \bibinfo {author} {\bibfnamefont {I.}~\bibnamefont
  {Brevik}},\ }\href@noop {} {\bibfield  {journal} {\bibinfo  {journal} {JOSA
  B}\ }\textbf {\bibinfo {volume} {12}},\ \bibinfo {pages} {2429} (\bibinfo
  {year} {1995})}\BibitemShut {NoStop}\bibitem [{\citenamefont {Almaas}\ and\ \citenamefont
  {Brevik}(2013)}]{Almaas2013}  \BibitemOpen
  \bibfield  {author} {\bibinfo {author} {\bibfnamefont {E.}~\bibnamefont
  {Almaas}}\ and\ \bibinfo {author} {\bibfnamefont {I.}~\bibnamefont
  {Brevik}},\ }\href {\doibase 10.1103/PhysRevA.87.063826} {\bibfield
  {journal} {\bibinfo  {journal} {Phys. Rev. A}\ }\textbf {\bibinfo {volume}
  {87}},\ \bibinfo {pages} {063826} (\bibinfo {year} {2013})}\BibitemShut
  {NoStop}\bibitem [{\citenamefont {Wang}\ and\ \citenamefont {Chan}(2014)}]{Wang2014}  \BibitemOpen
  \bibfield  {author} {\bibinfo {author} {\bibfnamefont {S.}~\bibnamefont
  {Wang}}\ and\ \bibinfo {author} {\bibfnamefont {C.}~\bibnamefont {Chan}},\
  }\href@noop {} {\bibfield  {journal} {\bibinfo  {journal} {Nat. Commun.}\
  }\textbf {\bibinfo {volume} {5}},\ \bibinfo {pages} {3307} (\bibinfo {year}
  {2014})}\BibitemShut {NoStop}\bibitem [{\citenamefont {Canaguier-Durand}\ and\ \citenamefont
  {Genet}(2014)}]{Canaguier-Durand2014}  \BibitemOpen
  \bibfield  {author} {\bibinfo {author} {\bibfnamefont {A.}~\bibnamefont
  {Canaguier-Durand}}\ and\ \bibinfo {author} {\bibfnamefont {C.}~\bibnamefont
  {Genet}},\ }\href@noop {} {\bibfield  {journal} {\bibinfo  {journal} {Phys.
  Rev. A}\ }\textbf {\bibinfo {volume} {89}},\ \bibinfo {pages} {033841}
  (\bibinfo {year} {2014})}\BibitemShut {NoStop}\bibitem [{\citenamefont {Antognozzi}\ \emph {et~al.}(2016)\citenamefont
  {Antognozzi}, \citenamefont {Bermingham}, \citenamefont {Harniman},
  \citenamefont {Simpson}, \citenamefont {Senior}, \citenamefont {Hayward},
  \citenamefont {Hoerber}, \citenamefont {Dennis}, \citenamefont {Bekshaev},
  \citenamefont {Bliokh},\ and\ \citenamefont {Nori}}]{Antognozzi2016}  \BibitemOpen
  \bibfield  {author} {\bibinfo {author} {\bibfnamefont {M.}~\bibnamefont
  {Antognozzi}}, \bibinfo {author} {\bibfnamefont {C.~R.}\ \bibnamefont
  {Bermingham}}, \bibinfo {author} {\bibfnamefont {R.~L.}\ \bibnamefont
  {Harniman}}, \bibinfo {author} {\bibfnamefont {S.}~\bibnamefont {Simpson}},
  \bibinfo {author} {\bibfnamefont {J.}~\bibnamefont {Senior}}, \bibinfo
  {author} {\bibfnamefont {R.}~\bibnamefont {Hayward}}, \bibinfo {author}
  {\bibfnamefont {H.}~\bibnamefont {Hoerber}}, \bibinfo {author} {\bibfnamefont
  {M.~R.}\ \bibnamefont {Dennis}}, \bibinfo {author} {\bibfnamefont {A.~Y.}\
  \bibnamefont {Bekshaev}}, \bibinfo {author} {\bibfnamefont {K.~Y.}\
  \bibnamefont {Bliokh}}, \ and\ \bibinfo {author} {\bibfnamefont
  {F.}~\bibnamefont {Nori}},\ }\href {http://dx.doi.org/10.1038/nphys3732}
  {\bibfield  {journal} {\bibinfo  {journal} {Nat Phys}\ }\textbf {\bibinfo
  {volume} {12}},\ \bibinfo {pages} {731} (\bibinfo {year} {2016})}\BibitemShut
  {NoStop}\bibitem [{\citenamefont {Zambrana-Puyalto}\ \emph {et~al.}(2013)\citenamefont
  {Zambrana-Puyalto}, \citenamefont {Fernandez-Corbaton}, \citenamefont {Juan},
  \citenamefont {Vidal},\ and\ \citenamefont {Molina-Terriza}}]{Zambrana2013}  \BibitemOpen
  \bibfield  {author} {\bibinfo {author} {\bibfnamefont {X.}~\bibnamefont
  {Zambrana-Puyalto}}, \bibinfo {author} {\bibfnamefont {I.}~\bibnamefont
  {Fernandez-Corbaton}}, \bibinfo {author} {\bibfnamefont {M.~L.}\ \bibnamefont
  {Juan}}, \bibinfo {author} {\bibfnamefont {X.}~\bibnamefont {Vidal}}, \ and\
  \bibinfo {author} {\bibfnamefont {G.}~\bibnamefont {Molina-Terriza}},\ }\href
  {\doibase 10.1364/OL.38.001857} {\bibfield  {journal} {\bibinfo  {journal}
  {Opt. Lett.}\ }\textbf {\bibinfo {volume} {38}},\ \bibinfo {pages} {1857}
  (\bibinfo {year} {2013})}\BibitemShut {NoStop}\bibitem [{\citenamefont {Bohren}\ and\ \citenamefont
  {Huffman}(1983)}]{Bohren1983}  \BibitemOpen
  \bibfield  {author} {\bibinfo {author} {\bibfnamefont {C.~F.}\ \bibnamefont
  {Bohren}}\ and\ \bibinfo {author} {\bibfnamefont {D.~R.}\ \bibnamefont
  {Huffman}},\ }\href {http://books.google.fr/books?id=S1RCZ8BjgN0C} {\emph
  {\bibinfo {title} {Absorption and scattering of light by small particles}}},\
  edited by\ \bibinfo {editor} {\bibnamefont {Wiley}},\ Wiley science paperback
  series\ (\bibinfo  {publisher} {Wiley},\ \bibinfo {address} {New York},\
  \bibinfo {year} {1983})\BibitemShut {NoStop}\bibitem [{\citenamefont {Zambrana-Puyalto}(2014)}]{ZambranaThesis}  \BibitemOpen
  \bibfield  {author} {\bibinfo {author} {\bibfnamefont {X.}~\bibnamefont
  {Zambrana-Puyalto}},\ }\emph {\bibinfo {title} {Control and characterization
  of nano-structures with the symmetries of light}},\ \href@noop {} {Ph.D.
  thesis},\ \bibinfo  {school} {Macquarie University} (\bibinfo {year}
  {2014})\BibitemShut {NoStop}\bibitem [{\citenamefont {Barton}\ \emph {et~al.}(1989)\citenamefont {Barton},
  \citenamefont {Alexander},\ and\ \citenamefont {Schaub}}]{Barton1989}  \BibitemOpen
  \bibfield  {author} {\bibinfo {author} {\bibfnamefont {J.}~\bibnamefont
  {Barton}}, \bibinfo {author} {\bibfnamefont {D.}~\bibnamefont {Alexander}}, \
  and\ \bibinfo {author} {\bibfnamefont {S.}~\bibnamefont {Schaub}},\
  }\href@noop {} {\bibfield  {journal} {\bibinfo  {journal} {J. Appl. Phys.}\
  }\textbf {\bibinfo {volume} {66}},\ \bibinfo {pages} {4594} (\bibinfo {year}
  {1989})}\BibitemShut {NoStop}\bibitem [{\citenamefont {Farsund}\ and\ \citenamefont
  {Felderhof}(1996)}]{Farsund1996}  \BibitemOpen
  \bibfield  {author} {\bibinfo {author} {\bibfnamefont {{\O}.}~\bibnamefont
  {Farsund}}\ and\ \bibinfo {author} {\bibfnamefont {B.}~\bibnamefont
  {Felderhof}},\ }\href@noop {} {\bibfield  {journal} {\bibinfo  {journal}
  {Physica A}\ }\textbf {\bibinfo {volume} {227}},\ \bibinfo {pages} {108}
  (\bibinfo {year} {1996})}\BibitemShut {NoStop}\bibitem [{\citenamefont {Nieminen}\ \emph {et~al.}(2014)\citenamefont
  {Nieminen}, \citenamefont {du~Preez-Wilkinson}, \citenamefont {Stilgoe},
  \citenamefont {Loke}, \citenamefont {Bui},\ and\ \citenamefont
  {Rubinsztein-Dunlop}}]{Timo2014}  \BibitemOpen
  \bibfield  {author} {\bibinfo {author} {\bibfnamefont {T.~A.}\ \bibnamefont
  {Nieminen}}, \bibinfo {author} {\bibfnamefont {N.}~\bibnamefont
  {du~Preez-Wilkinson}}, \bibinfo {author} {\bibfnamefont {A.~B.}\ \bibnamefont
  {Stilgoe}}, \bibinfo {author} {\bibfnamefont {V.~L.}\ \bibnamefont {Loke}},
  \bibinfo {author} {\bibfnamefont {A.~A.}\ \bibnamefont {Bui}}, \ and\
  \bibinfo {author} {\bibfnamefont {H.}~\bibnamefont {Rubinsztein-Dunlop}},\
  }\href@noop {} {\bibfield  {journal} {\bibinfo  {journal} {J. Quant.
  Spectrosc. Radiat. Transfer}\ }\textbf {\bibinfo {volume} {146}},\ \bibinfo
  {pages} {59} (\bibinfo {year} {2014})}\BibitemShut {NoStop}\bibitem [{\citenamefont {Palik}\ and\ \citenamefont
  {Ghosh}(1998)}]{Palik1998}  \BibitemOpen
  \bibfield  {author} {\bibinfo {author} {\bibfnamefont {E.}~\bibnamefont
  {Palik}}\ and\ \bibinfo {author} {\bibfnamefont {G.}~\bibnamefont {Ghosh}},\
  }\href {http://books.google.com/books?id=rxuG1kXvSqgC} {\emph {\bibinfo
  {title} {Handbook of optical constants of solids}}}\ (\bibinfo  {publisher}
  {Academic Press, Boston},\ \bibinfo {year} {1998})\BibitemShut {NoStop}\bibitem [{\citenamefont {Vuye}\ \emph {et~al.}(1993)\citenamefont {Vuye},
  \citenamefont {Fisson}, \citenamefont {Van}, \citenamefont {Wang},
  \citenamefont {Rivory},\ and\ \citenamefont {Abeles}}]{Vuye1993}  \BibitemOpen
  \bibfield  {author} {\bibinfo {author} {\bibfnamefont {G.}~\bibnamefont
  {Vuye}}, \bibinfo {author} {\bibfnamefont {S.}~\bibnamefont {Fisson}},
  \bibinfo {author} {\bibfnamefont {V.~N.}\ \bibnamefont {Van}}, \bibinfo
  {author} {\bibfnamefont {Y.}~\bibnamefont {Wang}}, \bibinfo {author}
  {\bibfnamefont {J.}~\bibnamefont {Rivory}}, \ and\ \bibinfo {author}
  {\bibfnamefont {F.}~\bibnamefont {Abeles}},\ }\href@noop {} {\bibfield
  {journal} {\bibinfo  {journal} {Thin Solid Films}\ }\textbf {\bibinfo
  {volume} {233}},\ \bibinfo {pages} {166} (\bibinfo {year}
  {1993})}\BibitemShut {NoStop}\bibitem [{\citenamefont {Garc\'{i}a-Etxarri}\ \emph
  {et~al.}(2011)\citenamefont {Garc\'{i}a-Etxarri}, \citenamefont
  {G\'{o}mez-Medina}, \citenamefont {Froufe-P\'{e}rez}, \citenamefont
  {L\'{o}pez}, \citenamefont {Chantada}, \citenamefont {Scheffold},
  \citenamefont {Aizpurua}, \citenamefont {Nieto-Vesperinas},\ and\
  \citenamefont {S\'{a}enz}}]{Aitzol2011}  \BibitemOpen
  \bibfield  {author} {\bibinfo {author} {\bibfnamefont {A.}~\bibnamefont
  {Garc\'{i}a-Etxarri}}, \bibinfo {author} {\bibfnamefont {R.}~\bibnamefont
  {G\'{o}mez-Medina}}, \bibinfo {author} {\bibfnamefont {L.~S.}\ \bibnamefont
  {Froufe-P\'{e}rez}}, \bibinfo {author} {\bibfnamefont {C.}~\bibnamefont
  {L\'{o}pez}}, \bibinfo {author} {\bibfnamefont {L.}~\bibnamefont {Chantada}},
  \bibinfo {author} {\bibfnamefont {F.}~\bibnamefont {Scheffold}}, \bibinfo
  {author} {\bibfnamefont {J.}~\bibnamefont {Aizpurua}}, \bibinfo {author}
  {\bibfnamefont {M.}~\bibnamefont {Nieto-Vesperinas}}, \ and\ \bibinfo
  {author} {\bibfnamefont {J.~J.}\ \bibnamefont {S\'{a}enz}},\ }\href@noop {}
  {\bibfield  {journal} {\bibinfo  {journal} {Opt. Express}\ }\textbf {\bibinfo
  {volume} {19}},\ \bibinfo {pages} {4815} (\bibinfo {year}
  {2011})}\BibitemShut {NoStop}\bibitem [{\citenamefont {Shi}\ \emph {et~al.}(2012)\citenamefont {Shi},
  \citenamefont {Tuzer}, \citenamefont {Fenollosa},\ and\ \citenamefont
  {Meseguer}}]{Shi2012}  \BibitemOpen
  \bibfield  {author} {\bibinfo {author} {\bibfnamefont {L.}~\bibnamefont
  {Shi}}, \bibinfo {author} {\bibfnamefont {T.~U.}\ \bibnamefont {Tuzer}},
  \bibinfo {author} {\bibfnamefont {R.}~\bibnamefont {Fenollosa}}, \ and\
  \bibinfo {author} {\bibfnamefont {F.}~\bibnamefont {Meseguer}},\ }\href@noop
  {} {\bibfield  {journal} {\bibinfo  {journal} {Adv. Mater}\ }\textbf
  {\bibinfo {volume} {24}},\ \bibinfo {pages} {5934} (\bibinfo {year}
  {2012})}\BibitemShut {NoStop}\bibitem [{\citenamefont {Zambrana-Puyalto}\ and\ \citenamefont
  {Bonod}(2015)}]{Zambrana2015}  \BibitemOpen
  \bibfield  {author} {\bibinfo {author} {\bibfnamefont {X.}~\bibnamefont
  {Zambrana-Puyalto}}\ and\ \bibinfo {author} {\bibfnamefont {N.}~\bibnamefont
  {Bonod}},\ }\href {\doibase 10.1103/PhysRevB.91.195422} {\bibfield  {journal}
  {\bibinfo  {journal} {Phys. Rev. B}\ }\textbf {\bibinfo {volume} {91}},\
  \bibinfo {pages} {195422} (\bibinfo {year} {2015})}\BibitemShut {NoStop}\bibitem [{\citenamefont {Muljarov}\ \emph {et~al.}(2014)\citenamefont
  {Muljarov}, \citenamefont {Doost},\ and\ \citenamefont
  {Langbein}}]{Muljarov2014}  \BibitemOpen
  \bibfield  {author} {\bibinfo {author} {\bibfnamefont {E.}~\bibnamefont
  {Muljarov}}, \bibinfo {author} {\bibfnamefont {M.}~\bibnamefont {Doost}}, \
  and\ \bibinfo {author} {\bibfnamefont {W.}~\bibnamefont {Langbein}},\
  }\href@noop {} {\bibfield  {journal} {\bibinfo  {journal} {arXiv preprint
  arXiv:1409.6877}\ } (\bibinfo {year} {2014})}\BibitemShut {NoStop}\bibitem [{\citenamefont {Marston}\ and\ \citenamefont
  {Crichton}(1984)}]{Marston1984}  \BibitemOpen
  \bibfield  {author} {\bibinfo {author} {\bibfnamefont {P.~L.}\ \bibnamefont
  {Marston}}\ and\ \bibinfo {author} {\bibfnamefont {J.~H.}\ \bibnamefont
  {Crichton}},\ }\href@noop {} {\bibfield  {journal} {\bibinfo  {journal}
  {Phys. Rev. A}\ }\textbf {\bibinfo {volume} {30}},\ \bibinfo {pages} {2508}
  (\bibinfo {year} {1984})}\BibitemShut {NoStop}\bibitem [{\citenamefont {Rahimzadegan}\ \emph {et~al.}(2016)\citenamefont
  {Rahimzadegan}, \citenamefont {Alaee}, \citenamefont {Fernandez-Corbaton},\
  and\ \citenamefont {Rockstuhl}}]{Aso2016}  \BibitemOpen
  \bibfield  {author} {\bibinfo {author} {\bibfnamefont {A.}~\bibnamefont
  {Rahimzadegan}}, \bibinfo {author} {\bibfnamefont {R.}~\bibnamefont {Alaee}},
  \bibinfo {author} {\bibfnamefont {I.}~\bibnamefont {Fernandez-Corbaton}}, \
  and\ \bibinfo {author} {\bibfnamefont {C.}~\bibnamefont {Rockstuhl}},\
  }\href@noop {} {\bibfield  {journal} {\bibinfo  {journal} {arXiv preprint
  arXiv:1605.03945}\ } (\bibinfo {year} {2016})}\BibitemShut {NoStop}\bibitem [{\citenamefont {M{\"a}kitalo}\ \emph {et~al.}(2014)\citenamefont
  {M{\"a}kitalo}, \citenamefont {Kauranen},\ and\ \citenamefont
  {Suuriniemi}}]{Makitalo2014}  \BibitemOpen
  \bibfield  {author} {\bibinfo {author} {\bibfnamefont {J.}~\bibnamefont
  {M{\"a}kitalo}}, \bibinfo {author} {\bibfnamefont {M.}~\bibnamefont
  {Kauranen}}, \ and\ \bibinfo {author} {\bibfnamefont {S.}~\bibnamefont
  {Suuriniemi}},\ }\href@noop {} {\bibfield  {journal} {\bibinfo  {journal}
  {Phys. Rev. B}\ }\textbf {\bibinfo {volume} {89}},\ \bibinfo {pages} {165429}
  (\bibinfo {year} {2014})}\BibitemShut {NoStop}\bibitem [{\citenamefont {Zambrana-Puyalto}\ \emph {et~al.}(2012)\citenamefont
  {Zambrana-Puyalto}, \citenamefont {Vidal},\ and\ \citenamefont
  {Molina-Terriza}}]{Zambrana2012}  \BibitemOpen
  \bibfield  {author} {\bibinfo {author} {\bibfnamefont {X.}~\bibnamefont
  {Zambrana-Puyalto}}, \bibinfo {author} {\bibfnamefont {X.}~\bibnamefont
  {Vidal}}, \ and\ \bibinfo {author} {\bibfnamefont {G.}~\bibnamefont
  {Molina-Terriza}},\ }\href {\doibase 10.1364/OE.20.024536} {\bibfield
  {journal} {\bibinfo  {journal} {Opt. Express}\ }\textbf {\bibinfo {volume}
  {20}},\ \bibinfo {pages} {24536} (\bibinfo {year} {2012})}\BibitemShut
  {NoStop}\end{thebibliography}
\end{document}